\documentclass[12pt]{article}
\usepackage{amsmath}
\usepackage{amsthm}
\usepackage{mathpazo}
\usepackage{setspace}
\usepackage{graphicx}
\usepackage{xcolor}
\usepackage{times}
\usepackage{comment}
\usepackage{placeins}
\usepackage{subcaption, booktabs}
\usepackage{hyperref}
\usepackage{authblk} 
\usepackage{array}
\usepackage{enumitem}
\usepackage{blkarray}
\newcolumntype{H}{>{\setbox0=\hbox\bgroup}c<{\egroup}@{}}
\graphicspath{{./ms and old stuff/plots/}}

\usepackage{float}
\usepackage{bm}
\usepackage{natbib}

\DeclareMathOperator{\Var}{\text{Var}}

\DeclareMathOperator{\Exp}{\text{Exp}}

\DeclareMathOperator{\Gr}{\text{Gr}}

\newcommand*{\bv}{\mathbf{v}}
 
\newcommand*{\bU}{\mathbf{U}}
\newcommand*{\bV}{\mathbf{V}}
\newcommand*{\bZ}{\mathbf{Z}}
\newcommand*{\bI}{\mathbf{I}}
\newcommand*{\bB}{\mathbf{B}}
\newcommand*{\bGamma}{\bm{\Gamma}}
\newcommand*{\bX}{\mathbf{X}}
\newcommand*{\bK}{\mathbf{K}}
\newcommand*{\by}{\bm{y}}
\newcommand*{\boldf}{\textbf{f}}
\newcommand*{\boldh}{\textbf{h}}
\newcommand*{\boldv}{\textbf{v}}
\newcommand*{\bF}{\mathbf{F}}
\newcommand*{\bC}{\mathbf{C}}

\newcommand*{\bM}{\mathbf{M}}

\newcommand*{\bH}{\mathbf{H}}
\newcommand*{\bL}{\mathbf{L}}
\newcommand*{\bPsi}{\bm\Psi}
\newcommand*{\bW}{\mathbf{W}}
\newcommand*{\bTheta}{\bm\Theta}
\newcommand*{\vech}{\text{vech}}
\newcommand{\T}{\top}
\newcommand{\bT}{\mathbf{T}}
\newcommand{\bmu}{\bm\mu}
\newcommand{\boldeta}{\bm\eta}
\newcommand{\bSigma}{\bm\Sigma}
\DeclareMathOperator*{\argmin}{arg\,min}

\newcommand*{\bA}{\textbf{A}}
\newcommand*{\GPFC}{\textrm{GPFC}}
\newcommand*{\SPFC}{\textrm{SPFC}}

\newtheorem{theorem}{Theorem}
\newtheorem{proposition}{Proposition}
\usepackage[parfill]{parskip}

\makeatletter
\g@addto@macro\normalsize{%
  \setlength\abovedisplayskip{6pt}
  \setlength\belowdisplayskip{6pt}
  \setlength\abovedisplayshortskip{6pt}
  \setlength\belowdisplayshortskip{6pt}
}
\makeatother

\newtheorem{lemma}{Lemma}
\newtheorem{corollary}{Corollary}
\newtheorem{condition}{Condition}
\numberwithin{theorem}{section}
\numberwithin{lemma}{section}
\numberwithin{proposition}{section}
\numberwithin{corollary}{section}
\renewcommand{\thecondition}{C\arabic{condition}}

\newcommand{\revise}[1]{{\leavevmode\color{black}#1}}

\newcommand{\blind}{1}

\begin{document}
\def\spacingset#1{\renewcommand{\baselinestretch}%
{#1}\small\normalsize} \spacingset{1}


\if1\blind
{
  \title{\bf \revise{Random effects model-based sufficient dimension reduction for independent clustered data}}
  \author{Linh H. Nghiem\thanks{Thanks to Andrew Wood and Janice Scealy for useful discussions.}\\
    School of Mathematics and Statistics, University of Sydney, Australia\\
    Francis K.C. Hui \thanks{FKCH was supported by an Australia Research Council Discovery Project DP230101908.}\\
    Research School of Finance, Actuarial Studies and Statistics, The Australian National University, Canberra, Australia}
    \date{}
  \maketitle
} \fi

\if0\blind
{
  \bigskip
  \bigskip
  \bigskip
  \begin{center}
    {\Large\bf Random effects model-based sufficient dimension reduction \\[.5em] for independent clustered data}
\end{center}
  \medskip
} \fi

\bigskip


\begin{abstract}
Sufficient dimension reduction (SDR) is a popular class of regression methods which aim to find a small number of linear combinations of covariates that capture all the information of the responses i.e., a central subspace. The majority of current methods for SDR focus on the setting of independent observations, while the few techniques that have been developed for clustered data assume the linear transformation is identical across clusters.
In this article, we introduce random effects SDR, where cluster-specific random effect central subspaces are assumed to follow a distribution on the Grassmann manifold, and the random effects distribution is characterized by a covariance matrix that captures the heterogeneity between clusters in the SDR process itself. 
\revise{
We incorporate random effect SDR within a model-based inverse regression framework. Specifically, we propose a random effects principal fitted components model, where a two-stage algorithm is used to estimate the overall fixed effect central subspace, and predict the cluster-specific random effect central subspaces. We demonstrate the consistency of the proposed estimators, while simulation studies demonstrate the superior performance of the proposed approach compared to global and cluster-specific SDR approaches. We also present extensions of the above model to handle mixed predictors, demonstrating how random effects SDR can be achieved in the case of mixed continuous and binary covariates. Applying the proposed methods to study the longitudinal association between the life expectancy of women and socioeconomic variables across 117 countries, we find log income per capita, infant mortality, and income inequality are the main drivers of a two-dimensional fixed effect central subspace, although there is considerable heterogeneity in how the country-specific central subspaces are driven by the predictors.
}

\end{abstract}
\noindent%
\vfill

\newpage
\spacingset{1.9} 
  
\section{Introduction} \label{sec:intro}
Sufficient dimension reduction \citep[SDR,][]{ma2013review,li2018sufficient} is a class of statistical methods that assume the outcome depends on covariates via a small number of their linear combinations. These linear combinations are known as sufficient predictors and retain the full regression information between the response and all the covariates, thereby overcoming the curse of dimensionality. 
Since the pioneering work of 
\citet{li1991sliced}, a vast literature has developed on different approaches to SDR, from inverse-moment-based and regression-based methods \citep[e.g.,][]{li1991sliced,cook2008}, forward regression \citep[e.g.,][]{xia2002adaptive}, to semiparametric techniques \citep{ma2012semiparametric}.
Much research has also been done to combine SDR with various aspects of statistical inference e.g., penalized SDR for high-dimensional sparse dimension reduction \citep{lin2018consistency, nghiem2021sparse}, and SDR in the presence of error-prone covariates \citep{chen2022sufficient,nghiem2024likelihood}.

The vast majority of the SDR methods focus on the setting of independent observations. By contrast, the adaptation of SDR to clustered data settings remains relatively underdeveloped, even though such data are common found in disciplines such as medical, social, and ecological and environmental studies \citep{verbeke2009linear}.
\revise{For independent clustered or longitudinal data settings,} the most common method of statistical analysis involves fitting mixed effects models or variations thereof \citep{verbeke2009linear, fitzmaurice2012applied}, which combine fixed effects that are identical across clusters with random effects representing cluster-specific deviations away from the overall fixed effects. The random effects are assumed to come from some common (typically normal) distribution with a zero mean/location vector and a covariance matrix, the latter of which characterizes the degree of heterogeneity across clusters. 
Mixed effect models can thus be seen as a balance between a global fixed effects model that ignores all clustering in the data, and a cluster-specific fixed effects model that ignores possible shared information across clusters.    

\revise{Among the few research that have been done on SDR for independent clustered data, all have effectively assumed a global fixed effects model approach to dimension reduction, i.e., the direction of the linear transformation is identical across all the clusters.} For example, \citet{bi2015sufficient} and \citet{xu2016estimating} employed a marginal estimation equation approach where working correlation matrices were included to account for temporal correlations arising within clusters for longitudinal data, while \citet{hui2022sufficient} proposed a finite mixture approach where the mixture proportions are modeled as known function of sufficient predictors and random effects are added to the mixture means to account for within cluster correlations. Global fixed effects SDR is also assumed in \citet{pfeiffer2021least} and \citet{song2023structured}, who developed methods for matrix-valued predictors formed from the collection of all repeated measurements of covariates corresponding to each cluster. \revise{We also acknowledge the connected literature on single index models and variations thereof for independent clustered and longitudinal data \citep[e.g.,][]{pang2012estimation, tian2023multivariate}; all of these works again assume a global fixed effects model approach to dimension reduction}.
Such an assumption of the same direction for the linear transformation across all clusters may be restrictive, as it does not allow for heterogeneity between clusters when it comes to the SDR process itself. 

In this article, \revise{we introduce the idea of random effects sufficient dimension reduction for independent clustered data}, where heterogeneity across clusters of the dimension reduction operation is induced by assuming that the linear transformation of the covariates for each cluster is drawn from a common distribution. \revise{This leads to an overall, fixed effects sufficient dimension reduction, and cluster-specific random effects sufficient dimension reduction representing deviations away from this.} Like other mixed models, we characterize the degree of heterogeneity in the sufficient dimension operation by a random effects covariance matrix.
One immediate challenge in defining random effects SDR is that the direction of the linear transformation of the covariates is not unique, since it is invariant to any orthogonal rotation. That is, similar to other SDR techniques, the estimation target is the subspace spanned by the columns of the linear transformation for each cluster, which in our setting is both the overall fixed effect central subspace and the cluster-specific random effect central subspaces. Because all these central subspaces are elements on a Grassmann manifold, then one approach would be to define the random effects distribution on this manifold. However, sufficiently flexible distributions on this manifold typically contain intractable normalizing constants, and their parameters are not easily interpretable \citep[e.g.,][]{ scealy2019scaled, scealy2022score}.

To overcome the above challenge, \revise{we propose to construct the distribution of cluster-specific random effect central subspaces as the image of an exponential mapping of a distribution defined on a tangent space of the Grassmann manifold \citep{srivastava2016functional} at an overall fixed effect central subspace; we will define it formally in Section \ref{sec:randomeffectsCS}.}
This modeling approach has the advantage that the tangent subspace is a vector space, meaning we can assume a random effects distribution defined on the corresponding Euclidean space e.g., a matrix normal distribution with a covariance matrix characterizing the heterogeneity of sufficient predictors between clusters. 
\revise{
After defining a distribution for the cluster-specific central subspaces in this manner, we incorporate this into the framework of model-based inverse regression models for SDR, first via the principal fitted components (PFC) model \citep{cook2008} for continuous predictors, and then through exponential family inverse models \citep{bura2015sufficient,bura2023sufficient} to handle mixed predictors e.g., longitudinal data with both continuous and binary predictors. To our knowledge, this article is the first in the literature to establish such a concept of random effects sufficient dimension reduction, let alone embed in within the framework of model-based inverse regression. 

For estimation, we propose a two-stage algorithm  where we first use a global fixed effects SDR to estimate a parameter of the overall fixed effects subspace, and then apply the Monte-Carlo Expectation Maximization algorithm \citep{wei1990monte} to estimate the remaining parameters and predict the cluster-specific random effect central subspaces. Focusing on the random effects PFC model, we establish the consistency of the proposed estimators under general conditions when the number of clusters goes to infinity, while simulation studies show that our proposed random effects SDR methods perform strongly compared with global fixed effects inverse regression models that ignore the heterogeneity between clusters, and cluster-specific fixed effects inverse regression models that do not borrow strength across clusters. 
Finally, we apply the proposed model to perform random effects SDR on a longitudinal, mixed predictor dataset studying the relationship between female life expectancy and various socioeconomic variables across different countries. Results show that the fixed effect central subspace is driven mostly by
log income per capita, sex ratio and infant mortality, while children per woman and income inequality drive heterogeneity between the country-specific random effects central subspaces.}

\revise{
The remainder of this article is organized as follows. Section \ref{sec:randomeffectsCS} introduces the general concept of random effect central subspaces. Section \ref{sec:RPFC} proposes the random effects PFC (or RPFC) model for independent clustered data with continuous responses along with a two-step estimation procedure. The consistency of the proposed estimators for the RPFC model is established in Section \ref{sec:theory}, while Section \ref{sec:simulation} compares the proposed model with fixed effects SDR models via a simulation study. Section \ref{sec:RMIR} presents extensions of the random effects PFC model to handle mixed continuous and binary predictors, establishing the sufficient reduction in this cases, along with a second numerical study demonstrating its strong empirical performance. Section \ref{sec:dataapplication} applies the proposed model to perform random effects SDR on longitudinal socioeconomic data, while Section \ref{sec:conclusion} discusses some potential avenues for future research.
}

\section{Random effect central subspaces} 
\label{sec:randomeffectsCS}
Consider a set of $n$ independent clusters, such that for cluster $i=1,\ldots,n$ we let $y_{ij}$ denote the $j$th measurement of the response for $j=1,\ldots, m_i$, and $\bX_{ij}$ denote a corresponding vector of $p$ covariates. For each cluster, SDR implies the response only depends on the covariates via a small number of their linear combinations, 
\begin{equation}
y_{ij} \perp \bX_{ij} \mid \bX_{ij}^\top \bm\Gamma_i; \quad i = 1,\ldots,n; j = 1,\ldots, m_i,
\label{eq:SDR_cluster}
\end{equation}
\revise{
where $\bm\Gamma_i \in \mathbb{R}^{p\times d}$ with structural dimension $d < p$ for $i=1,\ldots, n$. Note we assume the same structural dimension $d$ for all clusters; see Section \ref{sec:conclusion} for a discussion on how could be extended. More importantly,}
model \eqref{eq:SDR_cluster} allows the directions $\bm\Gamma_i$ to potentially vary from cluster to another, with the idea being that the $\bm\Gamma_i$'s represent deviations from some overall direction $\bm\Gamma_0$, which we formally define later. 
Without loss of generality, assume each $\bm\Gamma_i$ is a semi-orthogonal matrix, $\bm\Gamma_i^\top \bm\Gamma_i = \mathbf{I}_{d}$ for $i=1,\ldots, n$. Similar to standard SDR, only the subspaces spanned by each $\bm\Gamma_i$, denoted here as $[\bGamma_i]$ (and analogously $[\bGamma_0]$) are unique and identifiable; all of these are points on the Grassmann manifold $\Gr(p,d)$.  

If the spaces $[\bm\Gamma_i]$ are identical for all $n$ clusters i.e., $[\bm\Gamma_i] = [\bm\Gamma_0]$ for all $i$, then there is no heterogeneity between clusters in terms of the sufficient dimension reduction process. As reviewed in Section \ref{sec:intro}, \revise{much of the current literature on SDR for independent clustered and longitudinal data has been developed under such an assumption.} On the other hand, if the central subspaces $[\bm\Gamma_i]$ are completely different from another, then in practice one could identify the cluster-specific central subspace separately with no information being shared across clusters. In this case, 
there is no borrowing of strength across clusters in the dimension reduction operation.

\revise{As a balance between the above two situations then, and analogous to other mixed models for clustered and longitudinal data, we propose that the cluster-specific central subspaces $[\bm\Gamma_i]$ follow a common distribution on the Grassmann manifold. Specifically, 
we assume all $[\bm\Gamma_i]$ are obtained from an exponential mapping of an overall fixed effect central subspace $[\bm\Gamma_0]$, via a random velocity vector $\bV_i$ on the tangent space $T_{[\bGamma_0]}\Gr(p,d)$ of the Grassman manifold at $[\bGamma_0]$. For the remainder of this article, we will refer to $[\bGamma_i]$ as the cluster-specific random effect central subspace for the $i$th cluster, and $[\bGamma_0]$ as the overall fixed effect central subspace.
A review of the tangent space and exponential mapping on a Grassmann manifold is provided in Supplemental Material S1, but to summarize we assume that
\begin{equation}
[\bm\Gamma_i] = \Exp_{[\bm\Gamma_0]}(\bV_i), \bV_i \in T_{[\bm\Gamma_0]}\Gr(p, d); \quad i = 1,\ldots,n, 
\label{eq:Gamma}
\end{equation}
where for a given $[\bU] \in \Gr(p, d)$, the exponential map $\operatorname{Exp}_{[\bU]}$ transfers a point $\bV \in T_{[\bU]} \Gr(p,d)$ to $[\widetilde{\bU}] \in \Gr(p,d)$. As an aside, note using the exponential mapping assumes the cluster-specific central subspaces $[\bm\Gamma_i]$ are not too far from the fixed effect central subspace $[\bGamma_0]$, specifically, $\text{dist}([\bGamma_0], [\bGamma_i]) < \pi/2$ where the distance measure is based on the Riemann metric.
A sufficient condition for this constraint is that
the smallest singular value of the $d\times d$ matrix $\bGamma_0^\top \bGamma_i$ is strictly greater than $\cos^{-1}\{\pi/(2d^{-1/2})\}$ e.g, for $d=1$ the two bases $\bGamma_0$ and $\bGamma_i$ are not orthogonal to each other; see Supplementary Material S1 for further detail. 
We regard such a condition as being consistent with the general notion that  for independent clustered data, the random effects represent cluster-specific deviations not too far from the overall fixed effects response. That is, the condition on the exponential mapping is consistent with the notion of borrowing shared information across clusters as part of the SDR operation, with all the $[\bGamma_i]$'s belong to a local neighborhood of $[\bGamma_0]$. 
} 

Next, we assume a random effects distribution for the $\mathbf{V}_i$'s, whose covariance matrix characterizes the variability of central subspace among the clusters. Since $\bV_i \in T_{[\bm\Gamma_0]}\Gr(p, d)$, then $\bV_i$ is orthogonal to $\bm\Gamma_0$ and so any multivariate distribution imposed on $\bV_i$ should only have a non-zero density in the complement subspace of $[\bm\Gamma_0]$. In this article, we examine one such distribution that satisfies these requirements, namely the singular matrix-valued normal (MN) distribution with zero mean vector and two covariance matrices $\bm\Sigma$ and $\bm\Omega$ that characterize the covariance among the rows and columns of $\bV_i$, respectively. Let, $\bV_i \sim \text{MN}_{p\times d}(\mathbf{0}, \bm\Sigma, \bm\Omega)$, where the dimensions of $\bm\Sigma$ and $\bm\Omega$ are $p\times p$ and $d\times d$, respectively, where the orthogonality between $\bV_i$ and $\bm\Gamma_0$ implies that $\bm\Sigma$ is also orthogonal to $\bm\Gamma_0$, and $\bm\Sigma$ has rank at most $p-d$. The joint density of $\bV_i$, defined on the subspace $\bV_i^\top \bm\Gamma_0 = \mathbf{0}$, is then given by
\begin{equation}
p(\bV_i; \bm{\Sigma}, \bm{\Omega})=\frac{\exp \left\{-\frac{1}{2} \operatorname{tr}\left(\bm\Omega^{-1}\mathbf{V}_i^\top \bm{\Sigma}^{-}\mathbf{V}_i\right)\right\}}{(2 \pi)^{pd / 2}|\bm{\Lambda}|^{1 / 2} |\bm{\Omega}|^{1 / 2}},  
\label{eq:singularnormaldensity}
\end{equation}
where $\bm\Lambda$ denotes an $(p-d) \times (p-d)$ diagonal matrix with elements containing the non-zero eigenvalues of $\bm\Sigma$, $\bm{\Sigma}^{-}$ denotes the Moore-Penrose inverse of $\bm\Sigma$, and $\operatorname{tr}(\cdot)$ and $\vert \cdot\vert$ denote the trace and  determinant operators, respectively. Based on the above formulation, we have the following result whose proof is given in Supplementary Material S2. 

\begin{lemma}
If the density function of $\bV_i$ is symmetric about zero, then $[E(\bm\Gamma_i)] = [\bm\Gamma_0]$, where the expectation is taken with regards to a uniform probability measure on $\mathbb{R}^{p\times d}$.
\label{lemma:1}
\end{lemma}

The above implies that by assuming a degenerate MN distribution, $[\bm\Gamma_0]$ is the mean direction across all $n$ clusters and aligns with its interpretation as the overall fixed effect central subspace. Note for any scalar $s>0$, we have $p(\bV_i; \bm\Sigma, \bm\Omega) = p(\bV_i; s\bm\Sigma, s^{-1}\bm\Omega)$. That is, the covariance matrices $\bm\Sigma$ and $\bm\Omega$ are only identifiable up to a scale. 
Without loss of generality then, we set $\bm\Omega$ to be a correlation matrix. Furthermore, in this article we will restrict $\bm\Omega = \mathbf{I}_d$ such that any two dimensions in the central subspace can vary independently from one another, and the variability of the central subspace among the clusters is solely characterized by $\bm\Sigma$. 
We leave the exploration of more complex structures of $\bm\Omega$ for future research. 

\begin{figure}[tb]
\centering
 \includegraphics[width=0.9\textwidth, trim={0 0 2cm 0}, clip]{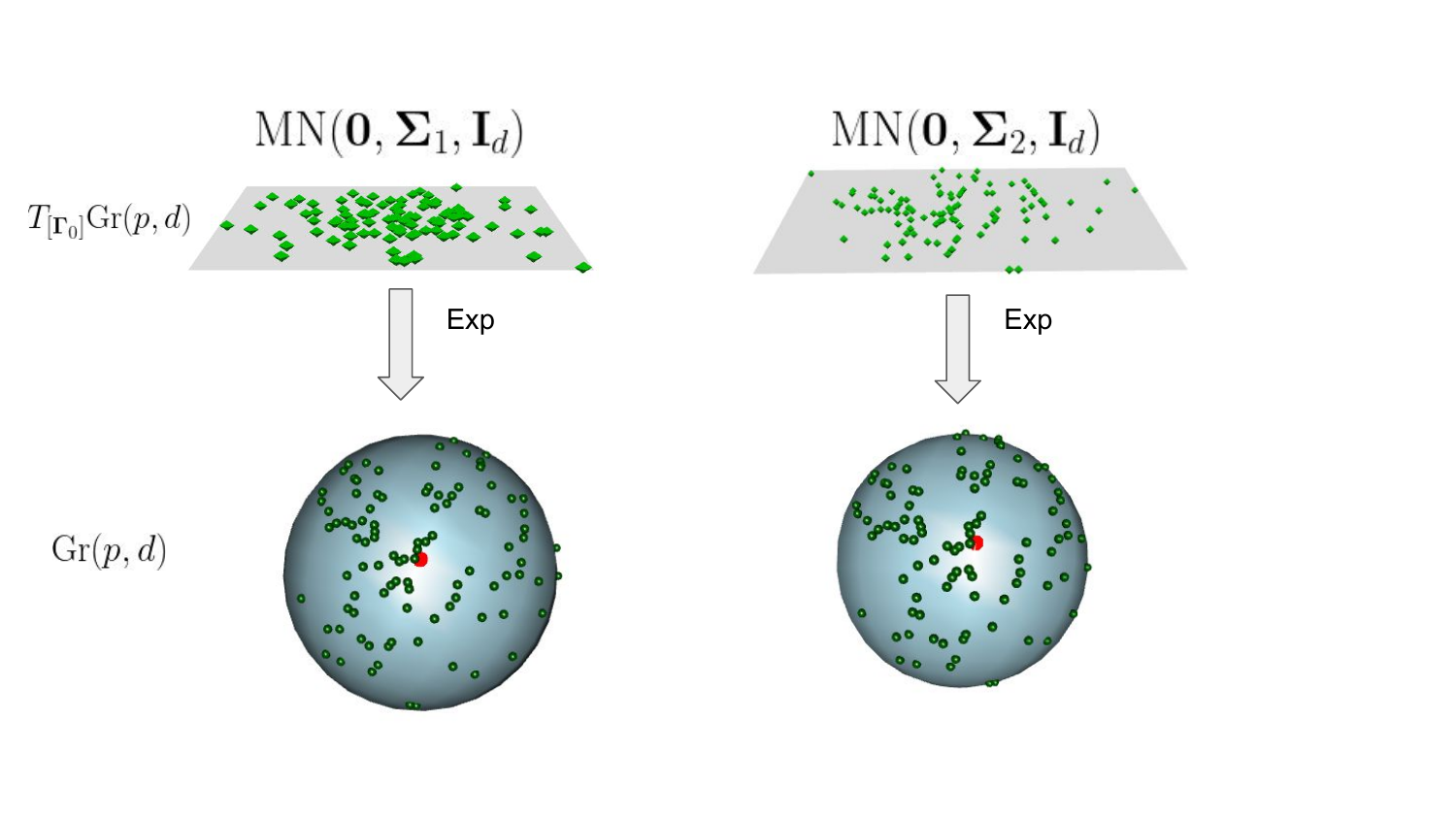}
\caption{
\revise{Visual illustration of cluster-specific random effect central subspaces based on the tangent space approach. In each panel, the overall fixed effect central subspace $[\bm\Gamma_0]$ is represented by the red dot, while the planes on the top row represent the tangent space $T_{[\bm\Gamma_0]}$. Each point on the tangent plane is given by $\bV_i \sim \text{MN}(\boldsymbol{0}, \bm\Sigma, \mathbf{I}_d)$, while each green point on the sphere below represents the corresponding $\bm\Gamma_i$ obtained via the exponential map in \eqref{eq:Gamma}. The left panel sets $\bm\Sigma_1 = \bK \widetilde{\bm\Sigma}_1 \bK$ with $\widetilde{\bm\Sigma}_1 = 0.3\bI_3$ and $\bK = \bI_p - \bGamma_0 \bGamma_0^\top$, while the right panel sets $\bm\Sigma_2 = \bK \widetilde{\bm\Sigma}_2 \bK$ where  $\widetilde{\bm\Sigma}_2$ has an exchangeable structure with variance 0.5 and covariance 0.20.
}
}
    \label{fig:demonstration}
\end{figure}

We conclude this section with a visual illustration of the cluster-specific random effect central subspaces based on the approach proposed above, for the simplest case with $d=1$ and $p=3$ i.e., a one-dimensional central subspace representing dimension reduction from three dimensions. In both panels of Figure \ref{fig:demonstration}, we set $\bm\Gamma_0 = (1, 0, 0)^\top$ and generate $n=50$ points (each one representing a cluster) $\bV_i = \mathbf{K}\widetilde{\bV}_i$, where $\mathbf{K}=\mathbf{I}_3 - \bm\Gamma_0 \bm\Gamma_0^\top$ and each $\widetilde{\bV}_i$ is sampled from the trivariate normal distribution $N_3(\mathbf{0}, \widetilde{\bm\Sigma})$. This data generation process ensures each $\bV_i$ is orthogonal to $\bm\Gamma_0$, and the corresponding covariance for $\bV_i$ is given by $\bm\Sigma = \bK \widetilde{\bm\Sigma} \bK$. The left and right panels then correspond to a particular choice for $\tilde{\bm\Sigma}$, namely an independence and an exchangeable structure. Finally, the corresponding $\bm\Gamma_i$ is obtained by applying equation \eqref{eq:Gamma} to each $\bV_i$. 
\revise{
From Figure \ref{fig:demonstration}, we see that although both panels have the same overall fixed effect central subspace, different covariances on the tangent space lead to different variability patterns in the manifold. That is, the covariance matrix $\bm\Sigma$ acts as a surrogate measure, in Euclidean space, for characterizing the heterogeneity among the cluster-specific random effect central subspaces. The left panel is based on a covariance matrix with equal variance values of 0.3 on the diagonal, where the corresponding (Monte-Carlo estimate of the) Fr\'{e}chet variance on the Grassmann manifold is 0.68 (the Fr\'{e}chet variance is an explicit quantification of the heterogeneity between clusters in terms of their Grassmann manifolds). Meanwhile, the right panel has slightly larger variance values of 0.5 on the diagonal, where the corresponding Fr\'{e}chet variance on the Grassmann manifold is 0.81. 

In the random effects SDR model formulations below, we treat $[\bm\Gamma_0]$ and $\bm\Sigma$ as unknown parameters which need to be estimated. 
}

\section{A random effects principal fitted components model}
\label{sec:RPFC}
\revise{We formulate an approach for performing random effects SDR by embedding the idea of cluster-specific random effect central subspaces within the model-based inverse regression framework. We start with the case of continuous predictors, and in Section \ref{sec:RMIR} we will extend this to the mixed predictor setting.}

Let $\{(\bX_{ij}, y_{ij}); i=1,\ldots, n, \; j=1,\ldots,m_i\}$ denote our set of observations, where $N = \sum_{i=1}^{n}m_i$ denotes the sample size and the $n$ clusters are assumed to be independent. Given an explicit parametric form for the random effects distribution of the cluster-specific central subspaces as defined in the preceding section, it is natural to adopt a model-based inverse-regression model for random effects SDR here. 
\revise{Compared with traditional sliced inverse regression and other inverse moments-based techniques \citep{li1991sliced}, a major advantage of using model-based inverse-regression methods is that they result in an explicit objective function for the parameters of the central subspace, and subsequently inherit desirable properties and practical tools from general likelihood theory e.g., for prediction and model selection as we will utilize later on in Sections \ref{subsec:estimation} and \ref{sec:simulation}. 
For independent clustered/longitudinal data specifically, model-based inverse-regression methods facilitate an appropriate conditional likelihood function of the cluster-specific central subspaces which, when combined with the ideas of Section \ref{sec:randomeffectsCS}, can be used to construct a formal marginal likelihood function based on integrating out random effect distribution.}

\revise{
With continuous predictors, and inspired by the work of \citet{cook2008}, we propose the random effect principal fitted components (RPFC) model as follows
\begin{subequations}
\begin{align} 
    \bX_{ij} &\mid (y_{ij}, \bm\Gamma_i) = \boldsymbol{\mu}_i + \bm\Gamma_i  \boldv_{ijy} + \bm\varepsilon_{ij} = \boldsymbol{\mu}_i + \bm\Gamma_i  \bC \boldf_{ijy} + \bm\varepsilon_{ij} , \quad \bm\varepsilon_{ij} \sim N_p(\boldsymbol{0}, \bm\Delta), \label{eqn:randomeffectsPFC_p1} \\
\bm\Gamma_i &= \Exp_{[\bm\Gamma_0]}(\bV_i), ~ \bV_i \stackrel{iid}{\sim} \text{MN}(\mathbf{0}, \bm\Sigma, \mathbf{I}_d), ~\bm\Sigma\bGamma_0 = \boldsymbol{0}, \quad i=1,\ldots, n, \label{eqn:randomeffectsPFC_p2}
\end{align}
\end{subequations}
where $\bm{\mu}_i \in \mathbb{R}^{p}$ denotes the conditional mean vector for the $i$th cluster, $\bm\Delta$ is a common $p\times p$ unstructured covariance matrix, and $\boldv_{ijy} \in \mathbb{R}^{d}$ is an unknown function of $y_{ij}$. In \eqref{eqn:randomeffectsPFC_p1}, the term $\bm\mu_i$ plays the role of cluster-specific intercept terms, and are treated as fixed parameters \citep[similar to fixed effect models in econometrics e.g.,][]{hsiao2002maximum}. To obtain the term on the right hand side of \eqref{eqn:randomeffectsPFC_p1}, we follow \citet{cook2008, cook2009dimension} 
and set $\boldv_{ijy} = \bC\left\{\boldf_{yij} - E\left(\boldf_{yij} \right) \right\}$, where $\bC \in \mathbb{R}^{d \times r}$ has rank $d \leq \min(p, r)$ and $\boldf_{ijy} \in \mathbb{R}^{r}$ is a known function of $y_{ij}$ often chosen to be a reasonably flexible set of basis functions of $y_{ij}$ e.g., piecewise polynomials. 
Without loss of generality, we assume that for each cluster $E(\boldf_{yij}) = \bm{0}$ for all $i=1,\ldots, n$; in practice, we center the basis functions within each cluster before conducting the analysis. Note also the parameters $\bC$ and $\bm\Delta$ in our formulation are constrained to be the same across clusters: while it is possible to vary them by cluster, empirically we found that doing so often leads to overfitting and general instability in estimation. Finally, the $\varepsilon_{ij}$'s are assumed to be independent of $\bm\Gamma_i$.

Turning to \eqref{eqn:randomeffectsPFC_p2}, using the distribution introduced in Section \ref{sec:randomeffectsCS}, we assume all the $\bm\Gamma_i$'s are images of an exponential map from a subspace $[\bm\Gamma_0]$ on the Grassmann manifold. With this formulation, Proposition \ref{prop:RPFCcentralsubspaces} establishes the forms of the cluster-specific and overall fixed effect central subspaces in the RPFC model.   

\begin{proposition}
\label{prop:RPFCcentralsubspaces}
For the RPFC model \eqref{eqn:randomeffectsPFC_p1}--\eqref{eqn:randomeffectsPFC_p2},  the cluster-specific central subspace for the $i$th cluster is given by $ [\bm\Theta_i] = \bm\Delta^{-1}[\bm\Gamma_i]$, and the overall fixed effect central subspace is given by $[\bm\Theta_0] = \bm\Delta^{-1}[\bm\Gamma_0]$.
\end{proposition}

The above results 
implies that within each cluster, there is no loss of information of $y_{ij}$ when transforming $p$-dimensional predictors $\bX_{ij}$ to $d<p$ dimensional sufficient predictors $\bGamma_i^\top\bm\Delta^{-1}\bX_{ij}$.  Furthermore, from equations~\eqref{eqn:randomeffectsPFC_p1}--\eqref{eqn:randomeffectsPFC_p2}, the marginal log-likelihood function of the RPFC model can be shown to be %
}
\begin{equation*}
\begin{aligned}
\ell(\bm\Gamma_0, \bm\mu_i, \bC, \bm\Delta, \bm\Sigma) & =  \sum_{i=1}^{n} \log \left[\int \left\{\prod_{j=1}^{m_i} \text{N}(\bX_{ij}; \bm\mu_i + \bm\Gamma_i \bm\beta \boldf_{ijy}, \bm\Delta)\right\} \text{MN}(\bV_i; \mathbf{0}, \bm\Sigma, \bI_d) d\bV_i \right] \\ & = \sum_{i=1}^{n} \log \ell_i(\bm\Gamma_0, \bm\mu_i, \bC, \bm\Delta, \bm\Sigma), 
\end{aligned}
\end{equation*}
where the integral is over all points on the tangent space $T_{[\bm\Gamma_0]}\Gr(d, p)$. 

\begin{proposition}
For any orthogonal matrix $\bA \in \mathbb{R}^{d \times d}$ such that $\bA \bA^\top = \bA^\top \bA = \mathbf{I}_d$, we have $\ell(\bm\Gamma_0, \bm\mu_i, \bC, \bm\Delta, \bm\Sigma) = \ell(\bm\Gamma_0 \bA, \bm\mu_i, ~\bA^\top\bC, \bm\Delta, \bm\Sigma)$.
\label{prop:1}
\end{proposition}

For the RPFC model then, the above proposition implies only the parameters $\bm\mu_i, \bm\Delta, \bm\Sigma$ are identifiable, while the  marginal likelihood is invariant to transformations of $\bm\Gamma_0$ and $\bC$. However, even though $\bm\Gamma_0$ is not identifiable its span $[\bm\Gamma_0]$ \emph{is} since it is invariant to any orthogonal rotation $\bm\Gamma_0 \bA$. In turn, the cluster-specific central subspaces $[\bm\Gamma_i]$ are also identifiable. 

\subsection{Two-stage estimation procedure} \label{subsec:estimation}
\revise{One challenge with fitting the RPFC model in equations~\eqref{eqn:randomeffectsPFC_p1}--\eqref{eqn:randomeffectsPFC_p2} is that while $\bGamma_0$ is not identifiable, the random effects covariance $\bm\Sigma$ is orthogonal to every vector belonging to $[\bm\Gamma_0]$. In other words, the parameter space for $\bm\Sigma$ depends upon the value of another parameter, posing issues with estimating $\bGamma_0$ and $\bm\Sigma$ jointly.
To overcome this, we develop a two-stage procedure where we first compute a consistent estimator of $[\bm\Gamma_0]$, and then we estimate the remaining parameters conditioned on this. Note the latter implies the estimated cluster-specific central subspaces will be in the local neighborhood of the estimated $[\hat\bGamma_0]$, consistent with the discussion below equation \eqref{eq:Gamma}. 
Throughout the developments, we assume the structural dimension $d$ is fixed and known; we address the issue of choosing $d$ via information criterion in Supplementary Material S4.
}

In the first stage, we estimate $[\bm\Gamma_0]$ by fitting a so-called global PFC model (GPFC) which ignores the clustering, $\bX_{ij} \mid y_{ij} = \widetilde{\bm\mu} + \widetilde{\bm\Gamma} \tilde{\bm\beta} \boldf_{ijy} + \tilde{\bm{\varepsilon}}_{ij},  ~\bm\varepsilon_{ij} \sim N_p(\boldsymbol{0}, \tilde{\bm\Delta})$. That is, we maximize the likelihood function
\begin{equation}
\ell_g(\widetilde{\bm\mu}, \widetilde{\bm\Gamma}, \widetilde{\bm\beta}, \widetilde{\bm\Delta}) = -\dfrac{N}{2} \log \vert \widetilde{\bm\Delta} \vert - \dfrac{1}{2} \sum_{i=1}^{n}\sum_{j=1}^{m_i}  \left(\bX_{ij} - \widetilde{\bm\mu} - \widetilde{\bm\Gamma} \widetilde{\bC} \boldf_{ijy} \right)^{\top} \widetilde{\bm\Delta}^{-1} \left(\bX_{ij} - \widetilde{\bm\mu} - \widetilde{\bm\Gamma} \widetilde{\bC} \boldf_{ijy} \right), 
\label{eq:globalPFCllh}
\end{equation}
the details of which can be found in \citet{cook2008}. Let the resulting maximum likelihood estimator be denoted by $\widehat{\bm\Gamma}_0$. 
We emphasize that GPFC only consistently estimates $[\bm\Gamma_0]$ and not the overall fixed effect central subspace $\bm\Delta^{-1}[\bm\Gamma_0]$. Moreover, equation \eqref{eq:globalPFCllh} does not offer predictions of the cluster-specific central subspaces. 

Given $\widehat{\bm\Gamma}_0$, then in the second stage of the estimation procedure we begin by removing the cluster-specific intercepts $\bm\mu_i$ (which are regarded as nuisance parameters) by subtracting each observation from its cluster mean in equation \eqref{eqn:randomeffectsPFC_p1}. This leads to a revised form of the RPFC model: $\bZ_{ij} = \bX_{ij} - \bar{\bX}_i = \bGamma_i\bC\boldh_{ijy} + \bm\epsilon_{ij}$ where $\bar{\bX}_i = m_i^{-1} \sum_{j=1}^{n} \bX_{ij}$, $\boldh_{ijy} = \boldf_{ijy} - m_i^{-1}\sum_{j=1}^{m_i}\boldf_{ijy}$, and the errors are defined as $\bm\epsilon_{ij} = \bm\varepsilon_{ij} - m_i^{-1} \sum_{j=1}^{m_i} \bm\varepsilon_{ij} \sim N_p(\bm{0}, (1-m_i^{-1}) \bm\Delta) $ for $i=1,\ldots, n$ and $j=1,\ldots, m_i$. Next, observe that $\sum_{j=1}^{m_i} \bZ_{ij} = \bm{0}$ and so it suffices to form the likelihood function based on $(m_i - 1)$ observations for each cluster. Moreover, 
the conditional covariance of any pair $(\bZ_{ij}, \bZ_{ij^\prime})$ is $-m_i^{-1} \bm\Delta$ for $j,j^\prime=1,\ldots, m_i$ and $j \neq j^\prime$. With this in mind, let $\by_i = (y_{i1}, \ldots, y_{im_i})^\top$ denote the full vector of responses for the $i$th cluster, and $\bZ_{i}$ be the $p \times (m_i-1)$ matrix whose $j$th column is given by $\bZ_{ij}$. Similarly, define $\bH_i$ as the $r \times (m_i-1)$ matrix whose $j$th column is given by $\boldh_{ijy}$. Conditional on $\bm{\Gamma}_i$ and $\by_i$ then, $\bZ_i$ follows a matrix normal distribution $\bZ_i \vert (\by_i, \bm\Gamma_i) \sim \text{MN}_{p\times (m_i-1)}(\bm\Gamma_i\bC\bH_i, \bm\Delta, \bL_i)$, where $\bL_i = \bI_{m_i-1} - m_i^{-1} \mathbf{J}_{m_i}$ where $\mathbf{J}_{m_i}$ is an $m_i \times m_i$ matrix of ones. The resulting complete data log-likelihood of the centered RFPC model, given $\widehat{\bm\Gamma}_0$, is then defined as
\begin{equation*}
\ell_c(\bm\Psi) = \sum_{i=1}^{n} \left[\log \left\{\text{MN}(\bZ_{i}; ~\bm\Gamma_i\bC\bH_i, \bm\Delta, \bL_i) \right\} + \log \left\{\text{MN}(\bV_i; \mathbf{0}, \bm\Sigma, \mathbf{I}_d)\right\}\right] = \sum_{i=1}^n \ell_{ci}(\bm\Psi), 
\end{equation*}
where $\widehat{\bm\Gamma}_0$ is implicit in the construction of each $\bm\Gamma_i$ i.e., $\bm\Gamma_i = \Exp_{[\widehat{\bm\Gamma}_0]}(\bV_i)$, and \\ $\bm\Psi = \{\bC^\top, \text{vech}(\bm\Delta)^\top, \text{vech}(\bm\Sigma)^\top\}^\top$ with $\text{vech}(\cdot)$ being the half-vectorization operator. 

\revise{
With the above setup, the second stage of the estimation procedure algorithm then uses a Monte-Carlo expectation-maximization \citep[MCEM,][]{wei1990monte} algorithm to estimate the remaining parameters. We provide details of this in Supplementary Material S3, but briefly we iterate (until convergence) between an E-step which involves using Monte-Carlo integration to compute the Q-function as $Q(\bm{\Psi}; \bm{\Psi}^{(0)}) = \sum_{i=1}^n \int \ell_{ci}(\bm{\Psi}) p(\bV_i;\bZ_{i}, \by_{i}, \bm{\Psi}^{(0)}) \, d \bV_i$, where $p(\bV_i; \bZ_{i}, \by_{i}, \bm{\Psi}^{(0)})$ generically denotes the conditional distribution of the random effects given the observed data and current estimates, and the M-step where we update the remaining parameters in the RPFC model as $\bm{\Psi}^{(1)} = \arg\max_{\bm{\Psi}} Q(\bm{\Psi}; \bm{\Psi}^{(0)})$ via a series of largely closed-form conditional updates. 
}

Let $\widehat{\bm\Psi} =  \{\widehat{\bC}^\top, \text{vech}(\widehat{\bm\Delta})^\top, \text{vech}(\widehat{\bm\Sigma})^\top\}^\top$  denote the estimator of $\bm\Psi$ upon convergence of the two-stage estimation procedure. Then the estimate of the overall fixed effect central subspace is given by $[\widehat{\bm\Theta}_0] = \widehat{\bm\Delta}^{-1}[\widehat{\bm\Gamma}_0]$. Furthermore, we can predict the cluster-specific random effect central subspaces as follows:
for the $i$th cluster, we first compute a prediction of $\mathbf{V}_i$ on the tangent space $T_{[\bm\Gamma_0]}\text{Gr}(p, d)$ as the mean of the conditional distribution $\widehat{\bV}_i = \int \mathbf{V}_i p(\mathbf{V}_i; \bX_{ij}, \bm{y}_{i}, \widehat{\bm{\Psi}})$; see Supplementary Material S3 for how this can be computed efficiency post-estimation. We then obtain a prediction as $\bm{\widehat\Gamma}_i = \Exp_{[\widehat{\bm\Gamma}_0]}(\widehat{\mathbf{V}}_i)$ and $[\widehat{\bm\Theta}_i] = \widehat{\bm\Delta}^{-1}[\widehat{\bm\Gamma}_i]$, for $i=1,\ldots, n$. 
While it is possible to use other predictors e.g., the mode or median of $p(\mathbf{V}_i; \bX_{ij}, \bm{y}_{i}, \widehat{\bm{\Psi}})$, in our empirical study we experimented with several choices and found that using the mean of the conditional distribution tended to be the most accurate.

\section{Asymptotic theory for the RPFC model}
\label{sec:theory}
In this section, we establish the consistency of the proposed 
estimators for the identifiable parameters in the RPFC model, in a setting where the number of clusters $n \to \infty$ and the cluster size $m_i$ is finite and bounded for all $i=1,\ldots, n$.  
Let $\widetilde{\bm\Psi}^* = (\bm\Gamma_0^{*\top}, \bC^{*\top}, \text{vech}(\bm\Delta^*)^{\top}, \text{vech}(\bm\Sigma^*)^{*\top})$ denote the true parameter value of $\bm\Gamma_0$, $\bC$, $\bm\Delta$ and $\bm\Sigma$ from \eqref{eqn:randomeffectsPFC_p1}--\eqref{eqn:randomeffectsPFC_p2}. We require the following regularity conditions.
\begin{condition}
$\widehat{\bm\Sigma}_{ff} = N^{-1}\sum_{i=1}^{n}\sum_{j=1}^     
     {m_i} \boldf_{ijy}\boldf_{ijy}^\top \to \bm{\Sigma}_{ff}$, where $\bm\Sigma_{ff}$  is  a $r\times r$ positive definite matrix, when $N \to \infty$.
\label{eq:condition1_GPFC}
\end{condition}

\begin{condition}
$\widehat{\bm\Sigma}_{xx} = N^{-1}\sum_{i=1}^{n}\sum_{j=1}^{m_i} \bX_{ij}\bX_{ij}^\top \stackrel{p}{\to} \bm\Sigma_{xx}$, where $\bm\Sigma_{xx}$  is  a $p\times p$ positive definite matrix when $N \to \infty$.
\label{eq:condition2_GPFC}
\end{condition}
Conditions 4.1 and 4.2 are mild since they essentially require $\bX_{ij}$ and $\boldf_{yij}$ to have finite (marginal) variances. We then have the following following result:
\begin{theorem}
Assume Conditions \ref{eq:condition1_GPFC} and \ref{eq:condition2_GPFC} are satisfied. Then $[\widehat{\bm\Gamma}_0]$ $\xrightarrow{p} [\bm\Gamma_0^*]$ as $N \rightarrow \infty$.
 \label{theorem:Gamma0}
\end{theorem}
Note Theorem \ref{theorem:Gamma0} only requires the total sample size $N \to \infty$, which is obviously satisfied when the number of clusters $n \to \infty$. It establishes the consistency of $[\widehat{\bm\Gamma}_{0}]$ obtained from fitting the GPFC model in the first stage of the estimation procedure. 

Next, we establish consistency of the estimated covariance matrices $\widehat{\bm\Sigma}$ and $\widehat{\bm\Delta}$ obtained from the second stage of estimation procedure. The latter is essential to guarantee consistency of the estimated overall fixed effect central subspace, while the former guarantees asymptotically correct estimation of the heterogeneity in the cluster-specific random effect central subspaces. Consistent with the above discussion, we will assume the true $\bm\Gamma_0^*$ is known up to an orthogonal rotation $\bm\Gamma_0^* \bA$. Consequently,
let $\bm\Psi^* = (\bA^\top\bC^*, \text{vech}(\bm\Delta^*)^\top, \text{vech}(\bm\Sigma^*)^\top)$ for any $d \times d$ orthogonal matrix  $\bA$ where $s = \dim(\bm\Psi^*) = \dim(\tilde{\bm{\Psi}}^*)$, and write the marginal log-likelihood function of the RPFC model at the second stage of estimation as $\ell(\bm\Psi \vert {\bGamma_0^* \bA}) = \sum_{i=1}^{n} \ell_i(\bm\Psi \vert {\bGamma_0^* \bA})$; see also Proposition \ref{prop:1}. Given the independence of the clusters, then without loss of generality, we assume $\ell_1(\bm\Psi \vert {\bGamma_0^* \bA})$ satisfies the following conditions:

\begin{condition}
The true parameter $\bm\Psi^*$ is an interior point of a compact parameter space, and $\ell_1(\bm\Psi \vert {\bGamma_0^* \bA})$ is distinct as a function of $\bm\Psi$. 
\label{condition1:RPFC}
\end{condition}

\begin{condition}
For all $\bm\Psi$ in an open set containing $\bm\Psi^*$, the $s\times s$ information matrix $\mathcal{I}(\bm{\Psi})$ with elements $\mathcal{I}_{jk}(\bm{\Psi}) = E\left( \partial^2\ell_1(\bm\Psi \vert {\bGamma_0^* \bA}) /\partial\Psi_j\partial\Psi_k \right); j,k=1,\ldots, s$ is positive definite with all its eigenvalues bounded away from zero and infinity.
\label{condition2:RPFC}
\end{condition}

\begin{condition}
For all $\bm\Psi$ in an open set containing $\bm\Psi^*$ and for $j,k, l=1,\ldots, s$, we have \\$\partial^3 \ell_1(\bm\Psi \vert {\bGamma_0^* \bA})/\partial\Psi_j\partial\Psi_k \partial\Psi_l$ exist and there exists functions $M_{ijk}(\bZ_1)$ such that \\ $\vert \partial^3 \ell_1(\bm\Psi \vert {\bGamma_0^* \bA})/\partial\Psi_j\partial\Psi_k \partial\Psi_l \vert \leq M_{ijk} (\bZ_1)$, where $E_{\bm\Psi^*}\{M_{ijk} (\bZ)\} < \infty$.
\label{condition3:RPFC}
\end{condition}

\revise{
The above are high-level conditions 
analogous to those often made when studying asymptotic properties for mixed models in general \citep[e.g.,][]{Nie2007,ibrahim2011fixed}.} Note in the SDR setting, since $\bm\Gamma_0^*$ is only identifiable up to an orthogonal rotation then we need to impose conditions on any such possible rotation. 

\begin{theorem}
Assume Conditions \ref{condition1:RPFC}--\ref{condition3:RPFC} holds. If the cluster size satisfy $m_i \geq 2$ for all $i=1,\ldots, n$ then $\widehat{\bm\Delta} \xrightarrow{p} \bm\Delta^*$ and $\widehat{\bm\Sigma} \xrightarrow{p} \bm\Sigma^*$ as $n \to \infty$. 
\label{thm:consistency}
\end{theorem}

The above result establishes the consistency of the two covariance matrices characterizing the overall fixed effect central subspace and the heterogeneity among cluster-specific random effect central subspaces, where the requirement for $m_i 
\ge$ 2 follows from the result of centering each observation around its cluster in the second stage of the estimation procedure.

\begin{corollary}
Assume Conditions \ref{eq:condition1_GPFC}-\ref{condition3:RPFC} hold. Then $\widehat{\bm\Delta}^{-1} [\widehat{\bm\Gamma}_0] \xrightarrow{p} \bm\Delta^{-1}[\bm\Gamma_0^*] $ when $n \to \infty$. 
\label{corrolary:fixedeffect}
\end{corollary}
 
This final result is a direct application of the Slutsky's theorem along with Theorems \ref{theorem:Gamma0} and \ref{thm:consistency}, and guarantees consistent estimation of the overall fixed effect central subspace.

\section{Simulation study} 
\label{sec:simulation}

We performed a numerical study to assess the performance of the RPFC model for random effects SDR with continuous predictors.
We simulated independent clustered data from two inverse models as follows: for $i=1,\ldots, n$ and $j=1, \ldots, m_i$, we first generated $y_{ij} \sim \text{N}(0, 1)$ and set $\bm\Delta$ to be an AR(1) correlation matrix with autocorrelation parameter 0.5. Next, we generated $\bm\Gamma_{i}$ following \eqref{eq:Gamma}, with $\bm\Gamma_0$ constructed from the QR decomposition of a random $p \times d$ matrix whose elements are generated from the uniform distribution between $(-1, 1)$, and the $p \times d$ matrices $\bV_i \sim \text{MN}(\boldsymbol{0}, \bm\Sigma, \mathbf{I}_d)$ where $\bm{\Sigma}= \bK \widetilde{\bm\Sigma} \bK$ and $\bK = \mathbf{I}_p - \bm\Gamma_0\bm\Gamma_0^\top$. 
Finally, for each cluster and conditional $\bV_i$ and $\bm{\Gamma}_0$, we simulated the covariates as

\begin{enumerate}
\item[(M1)] $\bX_{ij}  = \bm\Gamma_{i} v_{yij} + \bm{\varepsilon}_{ij}$, $v_{yij} = y_{ij} + (1/2) y_{ij}^2 + (1/3) y_{ij}^3$; \; $\bm\varepsilon_{ij} \sim N_p(\mathbf{0}, \bm\Delta)$,
\item[(M2)] $\bX_{ij} = \bm\Gamma_{i} \bv_{yij} + \bm{\varepsilon}_{ij}$,  $\bv_{yij} = [y_{ij} + (1/2) y_{ij}^2 + (1/3) y_{ij}^3, ~ y_{ij}]$; \; $\bm\varepsilon_{ij} \sim N_p(\mathbf{0}, \bm\Delta)$.
\end{enumerate}
Note that the structural dimensions are $d=1$ and $d=2$ for models M1 and M2, respectively, and are assumed to be known in this simulation design; see Supplementary Material S4 for developments in choosing $d$. For both models, we set $p = 7$ covariates and 
$\bm\Sigma = \bK \widetilde{\bm\Sigma} \bK$ with one of three possible structures of $\widetilde{\bm\Sigma}$: a diagonal form $ \widetilde{\bm\Sigma} = 0.5\mathbf{I}_p$, an AR(1) form with variance set to 0.3 and autocorrelation parameter set to $0.5$, and an exchangeable structure where all diagonal elements are set to 0.5 and all off-diagonal elements set to 0.1. 

For each combination of models M1 and M2 and the three choices of $\bm{\Sigma}$, the true overall fixed effects and cluster-specific random effect central subspaces are given by $[\bm\Theta_0] = \bm\Delta^{-1}[\bm\Gamma_0]$ and $[\bm\Theta_i] = \bm\Delta^{-1}[\bm\Gamma_i]$, respectively. We set the number of clusters to $n \in \{100, 500, 1000\}$, and for each $n$, we simulated the cluster sizes $m_i$ to be randomly integers between $10$ and $15$ inclusive. We generated 200 simulated datasets per simulation setting. 

For each simulated dataset, we compared the following three methods:
\begin{itemize}[leftmargin=*]
\item The RPFC model in \eqref{eqn:randomeffectsPFC_p1}--\eqref{eqn:randomeffectsPFC_p2} assuming the random effects covariance matrix to be unstructured. That is, no assumption about the structure $\bm\Sigma$ is made except the requirement $\bm\Sigma \bm\Gamma_0 = \boldsymbol{0}$. 

\item A global fixed effects PFC (GPFC) model, which ignores the clustered nature of the data and fits a single PFC model to the entire dataset.
The estimator for both the overall fixed effect central subspace and cluster-specific central subspaces is the same here, and given by $\widehat{\bm\Theta}_0^{\GPFC} = \widehat{\bm\Theta}_i^{\GPFC} =\left(\widehat{\bm\Delta}^{\GPFC}\right)^{-1} [\widehat{\bm\Gamma}_0]$, where $\widehat{\bm{\Gamma}}_0$ was introduced below equation \eqref{eq:globalPFCllh}. By construction, GPFC does not produce estimators of 
$\bm\Sigma$.
    
\item A separate fixed effects PFC (SPFC) model, which as the name suggests fits a separate PFC model to cluster. That 
is, for $i = 1,\ldots,n$ we maximize 
    \begin{align*} 
    \ell_i(\widetilde{\bm\mu}_i, \widetilde{\bm\Gamma}_i, \widetilde{\bC}_i, \widetilde{\bm\Delta}_i) &= -\frac{m_i}{2} \log \vert \widetilde{\bm\Delta}_i \vert - \frac{1}{2}\sum_{j=1}^{m_i} \left(\bX_{ij} - \widetilde{\bm\mu}_i - \widetilde{\bm\Gamma}_i\widetilde{\bC}_i \boldf_{ijy} \right)^{\top} \widetilde{\bm\Delta}_i^{-1} \left(\bX_{ij} - \widetilde{\bm\mu}_i - \widetilde{\bm\Gamma}_i \widetilde{\bC}_i \boldf_{ijy} \right).
    \end{align*}
    Let $\widehat{\bm\Delta}^{\SPFC}_i$ and $\widehat{\bm\Gamma}_i^{\SPFC}$ denote the resulting estimates for $\tilde{\bm\Delta}_i$ and $\tilde{\bm\Gamma}_i$, respectively. Then the cluster-specific central subspaces from SPFC are given by $[\widehat{\bm\Theta}_i^{\SPFC}] = \left(\widehat{\bm\Delta}^{\SPFC}_i\right)^{-1}[\widehat{\bm\Gamma}_i^{\SPFC}]$. Also, a reasonable estimator for the overall fixed effect central subspace $[\widehat{\bm{\Theta}}_0^\SPFC]$ is given by the sample Fr\'{e}chet mean of these estimates. 
    Finally, an estimate of $\bm\Sigma$ can be obtained by first performing an inverse exponential mapping  from $[\widehat{\bm{\Theta}}_i^\SPFC]$ to the tangent space of $\Gr(p, d)$ at $[\widehat{\bm{\Theta}}_0^\SPFC]$.
    Letting $\bm{\widetilde{V}}_i$ denote the image of that map, then we compute $\widehat{\bm\Sigma}^{\SPFC} = n^{-1} \sum_{i=1}^n \bm{\widetilde{V}}_i \bm{\widetilde{V}}_i^\top$.
\end{itemize}

For all three methods, we constructed $\boldf_{ijy}$ from polynomial bases with degree $r=4$ and centered them within each cluster such that $\sum_{j=1}^{m_i} \boldf_{ijy} = \boldsymbol{0}$. We assessed the performance of the RPFC, GPFC and SPFC models by the following three measures: 
(1) for estimating the overall fixed effect central subspace, we calculated the Frobenius norm of the difference between the projection matrix based on the estimate and the projection matrix based on the corresponding true value. That is, we computed $\Vert \widehat{\bm\Delta}^{-1}\mathcal{P}(\widehat{\bm\Gamma}_0)- {\bm\Delta}^{-1}\mathcal{P}(\bm\Gamma_0)\Vert_F$, where $\mathcal{P}(\bm{B}) = \bm{B}(\bm{B}^\top\bm{B})^{-1}\bm{B}$ for a generic matrix $\bm{B}$, and $\Vert \cdot \Vert_F$ denotes the Frobenius norm; 
(2) for estimating the random effects covariance matrix, we computed $\Vert \widehat{{\bm\Sigma}} - {\bm\Sigma} \Vert_F$; 
(3) for predicting the cluster-specific central subspaces, we computed the average Frobenius form across clusters, $n^{-1}\sum_{i=1}^n \Vert\mathcal{P}(\widehat{\bm\Theta}_i)- \mathcal{P}(\bm\Theta_i)\Vert_F$. 

Table \ref{tab:inversemodels-unstructured} demonstrates the RPFC model exhibits the best overall performance across the simulation settings considered. For estimating the overall fixed effect central subspace, SPFC produced the poorest performance, while even though RPFC and GPFC use the same estimator for $\bm\Gamma_0$ the former performs better due to its superior performance at estimating $\bm\Delta$. The estimation errors for both the overall fixed effect central subspace and random effect covariance matrix decreased noticeably when the number of clusters $n$ increased for RPFC, but not for SPFC. 
Turning to the estimation of the random effects covariance matrix, RPFC consistently outperformed SPFC.
Moreover, the estimation errors for both the overall fixed effect central subspace and random effect covariance matrix decreased noticeably when the number of clusters $n$ increased for RPFC, but not for SPFC. 
Finally, for predicting the cluster-specific central subspaces RPFC consistently outperformed SPFC, reflecting the ability of the former approach capacity to borrow strength across clusters in the SDR process.

\begin{table}[tb]
\centering
\caption{
Simulation results for random effects SDR with continuous predictors. The methods compared include the RPFC, GPFC, SPFC models. Performance is assessed in terms of estimating the fixed effect central subspace $\bm\Delta^{-1}[\bGamma_0]$, the random effects covariance matrix $\tilde{\bm\Sigma}$, and predicting the cluster-specific random effect central subspaces $[\bm\Theta_i]$. For each measure, the method/s with the lowest average Frobenius error in each row is highlighted.} 
\resizebox{\textwidth}{!}{\begin{tabular}{lrr  >{\bfseries}ccc  >{\bfseries}ccH  >{\bfseries}cc}
  \toprule[1.5pt]
   & & & \multicolumn{3}{c}{$\bm\Delta^{-1}[\bm\Gamma_0]$} & \multicolumn{3}{c}{${\bm\Sigma}$} & \multicolumn{2}{c}{$[\bm\Theta_i]$} \\ 
$\tilde{\bm\Sigma}$ & Model & $n$  & \normalfont{RPFC} & GPFC & SPFC & \normalfont{RPFC} & SPFC & GPFC & \normalfont{RPFC} & SPFC \\ 
\cmidrule(lr){4-6} \cmidrule(lr){7-9} \cmidrule(lr){10-11}
\addlinespace
Diagonal & M1 & 100 & 1.33 (0.49) & 1.35 (0.35) & 1.56 (0.32) & \normalfont 0.55 (0.08) & \bf 0.49 (0.02) & 0.79 (0.00) & 0.66 (0.03) & 1.20 (0.02)\\
 &  & 500 & 0.63 (0.22) & 1.11 (0.29) & 1.47 (0.26) & 0.43 (0.04) & 0.48 (0.01) & 0.79 (0.00) & 0.64 (0.02) & 1.20 (0.01)\\
 &  & 1000 & 0.44 (0.16) & 1.09 (0.30) & 1.43 (0.20) & 0.41 (0.03) & 0.47 (0.01) & 0.79 (0.00) & 0.63 (0.02) & 1.20 (0.01)\\
\addlinespace
 & M2 & 100 & 2.18 (0.39) & 2.05 (0.31) & 2.52 (0.32) & 0.71 (0.26) & 0.76 (0.09) & 0.79 (0.00) & 1.37 (0.03) & 1.59 (0.02)\\
 &  & 500 & 1.61 (0.47) & 1.83 (0.32) & 2.51 (0.32) & 0.53 (0.06) & 0.75 (0.07) & 0.79 (0.00) & 1.33 (0.03) & 1.59 (0.01)\\
 &  & 1000 & 1.27 (0.48) & 1.70 (0.29) & 2.48 (0.33) & 0.50 (0.05) & 0.74 (0.07) & 0.79 (0.00) & 1.31 (0.02) & 1.59 (0.01)\\
\addlinespace
AR(1) & M1 & 100 & 1.13 (0.43) & 1.36 (0.36) & 1.58 (0.29) & 0.74 (0.14) & 0.86 (0.03) & 0.98 (0.00) & 0.64 (0.06) & 1.24 (0.02)\\
 &  & 500 & 0.57 (0.23) & 1.23 (0.32) & 1.38 (0.25) & 0.65 (0.15) & 0.85 (0.02) & 0.98 (0.00) & 0.63 (0.05) & 1.23 (0.01)\\
 &  & 1000 & 0.39 (0.16) & 1.16 (0.31) & 1.33 (0.21) & 0.63 (0.14) & 0.85 (0.01) & 0.98 (0.00) & 0.63 (0.05) & 1.23 (0.01)\\
\addlinespace
 & M2 & 100 & 2.01 (0.46) & 2.09 (0.36) & 2.48 (0.34) & 0.90 (0.20) & 1.08 (0.11) & 0.98 (0.00) & 1.34 (0.04) & 1.63 (0.02)\\
 &  & 500 & 1.44 (0.49) & 1.83 (0.38) & 2.47 (0.36) & 0.77 (0.09) & 1.08 (0.10) & 0.98 (0.00) & 1.30 (0.05) & 1.62 (0.01)\\
 &  & 1000 & 1.13 (0.48) & 1.74 (0.34) & 2.48 (0.31) & 0.75 (0.09) & 1.07 (0.10) & 0.98 (0.00) & 1.28 (0.04) & 1.62 (0.01)\\
\addlinespace
Exchangeable & M1 & 100 & 0.94 (0.41) & 1.40 (0.36) & 1.52 (0.30) & 1.07 (0.49) & 1.47 (0.02) & 1.52 (0.00) & 0.59 (0.08) & 1.23 (0.02)\\
 &  & 500 & 0.43 (0.18) & 1.31 (0.32) & 1.30 (0.21) & 0.83 (0.33) & 1.46 (0.01) & 1.52 (0.00) & 0.56 (0.05) & 1.22 (0.02)\\
 &  & 1000 & 0.31 (0.12) & 1.30 (0.30) & 1.29 (0.19) & 0.77 (0.32) & 1.46 (0.01) & 1.52 (0.00) & 0.55 (0.04) & 1.23 (0.01)\\
\addlinespace
 & M2 & 100 & 1.80 (0.47) & 2.04 (0.36) & 2.45 (0.33) & 1.60 (0.61) & 1.62 (0.07) & 1.52 (0.00) & 1.25 (0.09) & 1.62 (0.02)\\
 &  & 500 & 1.23 (0.50) & 1.83 (0.40) & 2.45 (0.31) & 1.36 (0.38) & 1.62 (0.07) & 1.52 (0.00) & 1.17 (0.09) & 1.61 (0.02)\\
 &  & 1000 & 0.91 (0.40) & 1.75 (0.36) & 2.48 (0.33) & 1.32 (0.35) & 1.62 (0.07) & 1.52 (0.00) & 1.14 (0.09) & 1.60 (0.02)\\
 \bottomrule
   \label{tab:inversemodels-unstructured}
\end{tabular}}
\end{table}

\revise{
In Supplementary Material S3, we performed additional simulations where we set and assume in the fitting process an isotropic structure for $\bm\Sigma$, with results exhibiting similar trends to those above and the RPFC model performing best overall. Furthermore, in Supplementary Material S4 we propose and examine several information criteria that can be computed efficiently for selecting the structural dimension $d$ in the RPFC model. Empirical results suggest an AIC-type criterion constructed from an SPFC model performs reasonably well when the above simulation study was modified such that $d$ needed to be chosen from the candidate set $\{1,2,3,4,5\}$.}
\revise{
\section{Random effects SDR for mixed predictors}
\label{sec:RMIR}
The RPFC model proposed in Section \ref{sec:RPFC} is most relevant to situations where random effects SDR is applied to independent clustered/longitudinal datasets with only continuous covariates $\bX_{ij}$. However, in practice, longitudinal data often contain other types of covariates such as binary or count predictors.
In this section, we extend random effects SDR to the setting where, along with continuous predictors, we have $q \geq 1$ binary covariates which are either time-invariant or time-varying. We refer to the proposed extensions as random effects model-based inverse regression (RMIR) models, as they can be seen as generalizations of the RPFC model based on the exponential family inverse regression of \citet{bura2015sufficient} and \citet{bura2023sufficient}. For reasons of brevity, we focus the below developments on introducing two formulations of RMIR models, along with a simulation study to assess their performance; we leave details for their corresponding two-stage estimation procedures to Supplementary Material S5.


\subsection{Continuous and time-invariant binary predictors}\label{subsec:binarytimeinvariant}
Along with $p$-continuous covariates $\bX_{ij}$ and responses $y_{ij}$ for $i=1,\ldots,n$ and $j=1,\ldots, m_i$, suppose now we also record a set of $q$ time-invariant binary covariates $\bW_i  = (W_{i1}, \ldots, W_{iq})^\top \in \{0,1\}^q$. To be clear, by time-invariant, we mean those covariates that do not change with $j$ e.g., often gender in the context of longitudinal data. Let $\bm{y}_i = (y_{i1}, \ldots, y_{i, m_i})^\top$ and define $\bX_i$ as the $m_i \times p$ matrix of continuous covariates formed by stacking the $\bX_{ij}$'s as row vectors. 
Using the ideas of \citet{bura2023sufficient}, to achieve random effects SDR we propose to model the conditional joint distribution of $(\bX_i, \bW_i) \mid \bm{y}_i$ by decomposing it into a multivariate normal distribution for $\bX_i \mid (\bW_i, \bm{y}_i)$ and an Ising distribution model for $\bW_i \mid \bm{y}_i$. Coupling this with the ideas from Section \ref{sec:randomeffectsCS}, we have the following RMIR model: 
\begin{subequations}
\begin{align}
& \bX_{ij}\mid (y_{ij}, \bW_{i}, \bGamma_i) =\bm\mu + \bm\Gamma_i \bC_1 \boldf_{ijy} + \bm\beta (\bW_{i} - \bm\mu_W) + \bm{\varepsilon}_{ij}, ~ \bm{\varepsilon}_{ij} \sim N_p(\boldsymbol{0},\bm\Delta) \label{eqn:modelqbinary_p1} \\
& \bW_{i} \mid \bm{y}_{i} \propto \exp\left\{
\operatorname{vech}^\top(\bW_i\bW_i^\top) \operatorname{vech}^\top(\bm\tau_0 + \bB \bC_2 \bm{g}_{i\bm{y}})
\right\}, \label{eqn:modelqbinary_p2} \\
& \bm\Gamma_i = \Exp_{[\bm\Gamma_0]}(\bV_i), ~ \bV_i \stackrel{iid}{\sim} \text{MN}(\mathbf{0}, \bm\Sigma, \mathbf{I}_d), ~\bm\Sigma\bGamma_0 = \boldsymbol{0}; \quad i=1,\ldots,n. \label{eqn:modelqbinary_p3} 
\end{align}  
\end{subequations}
where in \eqref{eqn:modelqbinary_p1}, $\bm\mu\in \mathbb{R}^{p}$ denotes an overall intercept term for the continuous predictors, $\bm\mu_W  = \text{E}(\bm{W}_i)$, $\bm\beta \in \mathbb{R}^{p\times q}$, and $\bC_1 \in \mathbb{R}^{d \times r_1}$ and $\boldf_{yij} \in \mathbb{R}^{r_1}$ were (effectively) defined as part of the RPFC model in \eqref{eqn:randomeffectsPFC_p1}. Similar to the traditional mixed effect models, the time-invariant covariates $\bW_i$ are assumed only to have fixed effects in the conditional distribution $\bX_{ij}\mid (y_{ij}, \bW_{i}, \bGamma_i)$.
Turning to the Ising distribution model in \eqref{eqn:modelqbinary_p2}, $\bm{g}_{\bm{y}i}$ denotes a set of $r_2$ known basis functions, $\bm\tau_0 \in \mathbb{R}^{q(q+1)/2}$, $\bB \in \mathbb{R}^{q(q+1)/2\times d^\prime}$ and $\bC_2 \in \mathbb{R}^{d^\prime \times r_2}$, where $d^\prime$ is the structural dimension for the dimension reduction of $\by_i$ on $\bW_i$. One way to interpret such a model is as a set of logistic regressions for each component $W_{ik}$ conditional on other components $\bW_{i,-k} = (W_{i1}, \ldots, W_{i,k-1}, W_{i, k+1}, \ldots, W_{iq})^\top$, covariates $\bm{g}_{i\bm{y}}$,
\begin{equation*}
\log\left(\dfrac{P(W_{ik} = 1 \mid \bW_{i,-k}, \bm{y}_i)}{1 - P(W_{ik} = 1 \mid \bW_{i,-k}, \bm{y}_i)} \right) = \tau_{0kk} + \sum_{k \neq k^\prime} \tau_{0 k k^\prime} W_{i k^\prime} + \bm{b}_{kk}\bC_2\bm{g}_{i\bm{y}} + \sum_{k \neq k^\prime} \bm{b}_{k k^\prime} \bC_2\bm{g}_{i\bm{y}} W_{i k^\prime}, 
\end{equation*}
where $\tau_{0kk^\prime}$ and $\bm{b}_{kk^\prime}$ denote the row of $\bm\tau_0$ and $\bB$ corresponding to the pair $(k,k^\prime)$ respectively for $k,k^\prime = 1,\ldots, q$. Finally, \eqref{eqn:modelqbinary_p3} is the same as \eqref{eqn:randomeffectsPFC_p2} in the RPFC model for defining the cluster-specific random effect central subspaces. 

With this model, conditional on $\bGamma_i$ the joint distribution of $(\bX_i, \bW_i \mid \bm{y}_i)$ follows an exponential family inverse model. This leads the following result on the SDR operation of $\bm{y}_i$ on $(\bX_i, \bW_i)$. 

\begin{proposition}
For the RMIR model in \eqref{eqn:modelqbinary_p1}--\eqref{eqn:modelqbinary_p3}, the cluster-specific sufficient reduction for the regression of $\bm{y}_i$ on $(\bX_i, \bW_i)$ is given by $\widetilde{\bTheta}_i^\top   \mathbf{T}(\bX_{i}, \bW_{i})$, where
\begin{equation*}
\mathbf{T}(\bX_i, \bW_i) = \begin{pmatrix}
\operatorname{vec}(\bX_i^\top) \\
\bW_i \\ \operatorname{vech}\left(\bW_i \bW_i^\top\right)
\end{pmatrix}, ~\quad \widetilde{\bm{\Theta}}_i = \begin{pmatrix} 
\widetilde{\bm\Theta}_{i1} & \boldsymbol{0} \\
\boldsymbol{0} & \bB
\end{pmatrix}, ~ \widetilde{\bm\Theta}_{i1} =
\begin{pmatrix}
\bI_{m_i} \otimes \bm\Delta^{-1}\bGamma_i \\
- \boldsymbol{1}_{m_i}^\top \otimes \bm\beta^\top\bm\Delta^{-1}\bGamma_i
\end{pmatrix}.
\end{equation*}
The overall fixed effect central subspace corresponding to the continuous covariates is given by $\bm\Delta^{-1} [\bGamma_0]$, and the cluster-specific central subspace is given by $[\widetilde{\bm\Theta}_i]$.
\label{proposition:mixedcontinuous_qbinary}
\end{proposition}

The above result implies the cluster-specific central subspaces $[\bm\Theta_i]$ are separated into two parts:  $[\bm\Theta_{i1}]$ which exhibits heterogeneity across clusters, and $[\bB]$ which is common across all clusters. Moreover, the sufficient predictors for this RMIR model can be written as two components:  $[\bX_{ij}^\top \bm\Delta^{-1}\bGamma_i - \bW_i^\top \bm\beta^\top \bm\Delta^{-1}\bGamma_i, \vech(\bW_i\bW_i^\top)^\top\bB]$, where the first component acts on the observation level i.e., varies with $(i,j)$ while the second component acts only on the cluster level i.e., varies with $i$ only. This differs from the RPFC model in Section \ref{sec:RPFC} with all time-varying continuous predictors, where all sufficient predictors vary at the observation level. Furthermore, we note that although $\bW_i$ only has a fixed effect on the conditional distribution $\bX_{ij} \mid (y_{ij}, \bW_i, \bGamma_i)$, its effect of $\bW_i$ on the first component varies with the cluster-specific $\bGamma_i$. Indeed, the first component can also be more concisely written as $(\bX_{ij}^\top, \bW_i^\top) \bm\Theta_i$, where $\bm\Theta_i = (\bm\Gamma_i^\top\bm\Delta^{-1}, -\bm\Gamma_i^\top\bm\Delta^{-1}\bm\beta)^\top$; see also Corollary 5 in \citet{bura2023sufficient} for more discussion of this form of dimension reduction. 

\subsection{Continuous and time-varying binary predictors}
\label{subsec:binarytimevarying}
Suppose now the $q$ binary covariates are time-varying, such that $\bW_{ij} = (W_{ij1}, \ldots, W_{ijq})^\top$ for $i = 1,\ldots,n$ and $j = 1,\ldots,m_i$. 
Borrowing ideas from \citet{bura2015sufficient}, we present one example of random effects SDR where $W_{ijk} | y_{ij}$ is assumed to involve one random effect for $k = 1,\ldots,q$, but both fixed and random effects are incorporated into the conditional distribution $\bX_{ij} \mid (y_{ij}, \bW_{ij})$. Specifically, we formulate the following RMIR model:
\begin{subequations}
\begin{align}
&\bX_{ij}\mid (y_{ij}, \bW_{ij}) = \bm\mu + \bm\Gamma_i \bC_1\boldf_{yij}  + \bm\beta (\bW_{ij} - \bm\mu_W) + \bm\varepsilon_{ij}, ~ \bm\varepsilon_{ij} \sim N_p(\bm{0}, \bm\Delta) \label{eqn:model1binary_p1}   \\
& W_{ijk} \mid (y_{ij}, \bB_{ik}) \sim \text{Bern}(p_{ij}), ~\text{logit}(p_{ij}) = \bB_{ik}^\top \left\{\boldf_{yij} - E(\boldf_{yij})\right\}; \; \bB_{ik} \sim N(\bm{b}_{0k}, \bm\Sigma_{bk}) \label{eqn:model1binary_p2} \\
& \bm\Gamma_i = \Exp_{[\bm\Gamma_0]}(\bV_i), ~ \bV_i \stackrel{iid}{\sim} \text{MN}(\mathbf{0}, \bm\Sigma, \mathbf{I}_d), ~\bm\Sigma\bGamma_0 = \boldsymbol{0}; \quad i=1,\ldots,n, k=1,\ldots, q, \label{eqn:model1binary_p3}   
\end{align}
\end{subequations}
where $\bm{b}_{0k}, \bB_{ik} \in \mathbb{R}^{r}$, $\bm\Sigma_{bk}$ denotes the random effects covariance matrix for the conditional distribution of the $q$ binary predictor, $\text{logit}(\cdot)$ denotes the logit function, and all other notations are defined analogously as in Section \ref{subsec:binarytimeinvariant}. While \eqref{eqn:model1binary_p1} resembles the RMIR model in \eqref{eqn:modelqbinary_p1}, in \eqref{eqn:model1binary_p2} we model the $k$th binary covariate via a logistic mixed model with the pre-specified bases $\boldf_{yij}$, and assume the cluster-specific random slopes $\bB_{ik}$ are independent of each other. Finally, \eqref{eqn:model1binary_p3} characterizes the random effect central subspaces for the continuous predictors as we have done previously. Let $\bB_i \in \mathbb{R}^{q \times r}$ be the matrix formed by stacking all $\bB_{ik}$  together.  
Since all the covariates are time-varying, then it is natural that the random effects SDR operation acts on the observational level i.e., it varies with $(i,j)$, in an analogous manner to the RPFC model in Section \ref{sec:RPFC}. This idea is formalized below.
\begin{proposition}
\label{prop:1binary}
For the RMIR model in \eqref{eqn:model1binary_p1}--\eqref{eqn:model1binary_p3}, the cluster-specific sufficient reduction for the regression of $y_{ij}$ on $(\bX_{ij}, \bW_{ij}^\top)$ is given by $\bTheta_i^\top   \mathbf{T}(\bX_{ij}, \bW_{ij})$, where
\begin{equation*}  
\mathbf{T}(\bX_{ij}, \bW_{ij}) = \left(\bX_{ij}^\top, \bW_{ij}^\top \right)^\top, ~\quad \bm{\Theta}_i = \begin{bmatrix}
\bm\Delta^{-1}\bGamma_i \bC_1 \\ 
-\bm\beta^\top \bm\Delta^{-1}\bGamma_i\bC_1 + \bB_i
\end{bmatrix}.
\end{equation*}
The overall fixed effect central subspace corresponding to the continuous covariates is given by $\bm\Delta^{-1} [\bGamma_0]$, and the corresponding cluster-specific central subspace is given by $[\bm\Theta_i].$
\end{proposition}

In examining both the time-invariant and time-varying case, we have illustrated two ways in which random effects SDR with mixed continuous and binary covariates (via Ising and logistic mixed models) can be accomplished. Depending on the data and the precise goal of the random effects SDR, there may be other possibilities e.g., a mixture of time-varying and time-invariant binary predictors, or time-varying binary predictors pairwise interactions are involved in the random effects SDR; we leave these as avenues of future research.
 
\subsection{Simulation study for RMIR models}
\label{subsec:simulation_RMIR}

We simulated independent clustered data from both RMIR models  \eqref{eqn:modelqbinary_p1}--\eqref{eqn:modelqbinary_p3} and\eqref{eqn:model1binary_p1}--\eqref{eqn:model1binary_p3}, where for both we set $\boldf_{ijy} = y_{ij}$, $q=4$ binary predictors, and $\bC_1 = 3$. In the former, we set $\bm{\tau}_0 = \bm{0}$, $\bm{g}_{i \bm{y}} = m_i^{-1}\sum_{i=1}^{m_i} y_{ij}$, $d^\prime = 1$, and $\bB = (1, 1, \ldots, 1, 10, 10)/\sqrt{208}$ and $\bC_2 = 6$. That is, the SDR at the cluster level for the binary predictors \eqref{eqn:modelqbinary_p2} is dominated by the pairwise interaction terms $W_2W_4$ and $W_3 W_4$. In the latter, noting $r = 1$ with the setup above, we also set $\bm{b}_{0k} = 0.5$ and $\bm\Sigma_{bk} = \sigma_{bk}^2 = 0.6^2$ for all $k=1,\ldots, q$ for the random effects SDR in \eqref{eqn:model1binary_p2}. All other parameters for both RMIR models, along with the values of $n, m_i, \widetilde{\bm{\Sigma}}$ and the number of datasets for each combination, were constructed as in Section \ref{sec:simulation}. 

Analogous to Section \ref{sec:simulation}, for each simulated dataset we compared three methods for performing random effects SDR: 
(1) the corresponding RMIR model as formulated above;
(2) a global model-based inverse regression (GMIR) model of $\bX_{ij}$ on $\boldf_{ijy}$ and $\bW_{ij}$, which ignores all the clustering. This can be seen as an extension of the global PFC model; 
(3) a separate model-based inverse regression model (SMIR) for each cluster separately. This can be seen as an extension of the separate PFC model. Note since the binary covariate is time-invariant in Section \ref{subsec:binarytimeinvariant}, then we use the GMIR estimator of $\bm\beta$ in calculating the prediction of $[\bm\Theta_i]$ for SMIR.   

We assessed the performance similarly to Section \ref{sec:simulation} i.e., estimating the overall fixed effect central subspace corresponding to the continuous covariates $\bm\Delta^{-1} [\bGamma_0]$, the random effects covariance matrix $\bm\Sigma$, and predicting the component of the cluster-specific central subspaces $[\bm\Theta_i]$. 
Consistent with the empirical results for the RPFC model in Section \ref{sec:simulation}, RMIR exhibited the best performance among the three considered methods in estimating the overall fixed effect central subspace for continuous covariates in both models (Table \ref{tab:binaryresults}). For estimating the random effect covariance matrix $\bm\Sigma$,
RMIR produced lower errors than SMIR in the majority of settings, and this error tended to decrease when the number of clusters $n$ increased. Finally, for predicting the cluster-specific random effect central subspaces, RMIR also had consistently lower errors than SMIR.

\begin{table}[tb]
\centering
\caption{
\revise{
Simulation results for random effects SDR involving mixed continuous and time-invariant or time-varying binary predictors.
The methods compared include the proposed random effects model-based inverse regression (RMIR), global model-based inverse regression (GMIR), and separate model-based inverse regression (SMIR) models. Performance is assessed in terms of estimating $\bm\Delta^{-1}[\bGamma_0]$ for continuous covariates, the random effects covariance matrix $\tilde{\bm\Sigma}$, and predicting a component of the random effect central subspaces $[\bm\Theta_i]$. 
The method/s with the lowest average Frobenius error in each row are highlighted.} 
}
\resizebox{\textwidth}{!}{\begin{tabular}{l r  >{\bfseries}ccc  >{\bfseries}cc  >{\bfseries}cc}
  \toprule[1.5pt]
    & &   \multicolumn{3}{c}{$\bm\Delta^{-1}[\bm\Gamma_0]$} & \multicolumn{2}{c}{${\bm\Sigma}$} & \multicolumn{2}{c}{$[\bm\Theta_i]$} \\ 
$\tilde{\bm\Sigma}$ & $n$  & \normalfont{RMIR} & GMIR & SMIR & \normalfont{RMIR} & SMIR  & \normalfont{RMIR} & SMIR \\ 
\cmidrule(lr){3-5} \cmidrule(lr){6-7} \cmidrule(lr){8-9}
\addlinespace
\multicolumn{9}{c}{\emph{Time-invariant binary covariates}} \\
\addlinespace
Diagonal & 100 & \normalfont 1.73 (0.48) & \bf 1.63 (0.38) & 2.11 (0.57) & \normalfont 0.87 (0.16) & \bf 0.78 (0.05) & 0.42 (0.02) & 1.09 (0.06)\\
 & 500 & 1.18 (0.38) & 1.49 (0.37) & 2.02 (0.54) & 0.56 (0.10) & 0.75 (0.05) & 0.41 (0.01) & 1.10 (0.06)\\
 & 1000 & 0.91 (0.27) & 1.49 (0.36) & 1.99 (0.53) & 0.45 (0.07) & 0.74 (0.05) & 0.40 (0.01) & 1.11 (0.06)\\
\addlinespace
AR(1) & 100 & 1.73 (0.41) & 1.68 (0.35) & 1.71 (0.45) & 1.08 (0.23) & 1.13 (0.15) & 0.50 (0.04) & 1.12 (0.06)\\
 & 500 & 1.42 (0.39) & 1.64 (0.34) & 1.61 (0.50) & 0.86 (0.20) & 1.15 (0.17) & 0.47 (0.02) & 1.12 (0.06)\\
 & 1000 & 1.20 (0.42) & 1.58 (0.31) & 1.56 (0.45) & 0.71 (0.23) & 1.14 (0.15) & 0.47 (0.02) & 1.12 (0.05)\\
\addlinespace
Exchangeable & 100 & 1.80 (0.46) & 1.73 (0.38) & 2.03 (0.50) & \normalfont 1.01 (0.18) & \bf 0.99 (0.14) & 0.44 (0.03) & 1.11 (0.06)\\
 & 500 & 1.34 (0.42) & 1.59 (0.37) & 1.83 (0.51) & 0.68 (0.15) & 0.96 (0.16) & 0.42 (0.02) & 1.11 (0.06)\\
 & 1000 & 1.09 (0.38) & 1.53 (0.35) & 1.76 (0.44) & 0.60 (0.13) & 0.99 (0.16) & 0.43 (0.01) & 1.11 (0.06)\\
 \addlinespace
 \multicolumn{9}{c}{\emph{Time-varying binary covariates}} \\
\addlinespace
Diagonal & 100 & 1.81 (0.49) & 1.71 (0.39) & 2.70 (0.61) & \normalfont 0.90 (0.15) & \bf 0.79 (0.05) & 0.45 (0.03) & 1.07 (0.04)\\
 & 500 & 1.19 (0.39) & 1.53 (0.38) & 2.47 (0.65) & 0.56 (0.10) & 0.75 (0.05) & 0.44 (0.02) & 1.06 (0.03)\\
 & 1000 & 0.93 (0.28) & 1.43 (0.35) & 2.45 (0.71) & 0.47 (0.07) & 0.75 (0.06) & 0.44 (0.01) & 1.06 (0.03)\\
\addlinespace
AR(1) & 100 & 0.73 (0.24) & 1.37 (0.36) & 1.93 (0.52) & \normalfont 0.68 (0.41) & \bf 0.63 (0.14) & 0.53 (0.04) & 1.09 (0.04)\\
 & 500 & 0.46 (0.13) & 1.35 (0.35) & 1.80 (0.47) & 0.59 (0.38) & 0.63 (0.16) & 0.51 (0.03) & 1.09 (0.03)\\
 & 1000 & 0.38 (0.12) & 1.29 (0.38) & 1.68 (0.57) & 0.51 (0.38) & 0.62 (0.18) & 0.51 (0.03) & 1.09 (0.03)\\
\addlinespace
Exchangeable & 100 & 0.78 (0.25) & 1.39 (0.32) & 2.06 (0.54) & \normalfont 1.05 (1.21) & \bf 0.74 (0.19) & 0.50 (0.06) & 1.08 (0.04)\\
 & 500 & 0.48 (0.14) & 1.36 (0.36) & 1.92 (0.54) & 0.55 (0.37) & 0.77 (0.18) & 0.47 (0.03) & 1.08 (0.03)\\
 & 1000 & 0.41 (0.11) & 1.36 (0.32) & 1.93 (0.53) & 0.59 (0.38) & 0.77 (0.22) & 0.47 (0.03) & 1.08 (0.02)\\
 \bottomrule[1.5pt]
\end{tabular}}
\label{tab:binaryresults}
\end{table}

\section{Application to longitudinal socioeconomic data on female life expectancy}
\label{sec:dataapplication}
We illustrate an application of random effects SDR on socioeconomic data extracted from the Gapminder database. Briefly, the data contain multiple socioeconomic variables of $n=117$ countries collected in the years 1990--2015, and are publicly available at \\ \url{https://open-numbers.github.io/}. 
In this analysis, we focused on modeling the relationship between life expectancy of women across the countries ($y_{ij}$ for country $i$ and time point $j$) as a function of $p=6$ continuous predictors ($X_{ij1}$ = log income per capita; $X_{ij2}$ = sex ratio i.e., number of females per 100 males across all age groups; $X_{ij3}$ = infant mortality rate per 1000 new births; $X_{ij4}$ = emissions consumption per person; $X_{ij5}$ = the average children per woman; $X_{ij6}$ = income inequality via Gini index), and $q=2$ time-invariant binary covariates ($W_{i1}$ = whether the country belongs to the group of least developed countries (LDC) classified by the United Nations; $W_{i2}$ = whether the countries belong to Western (West) culture). Each country is assumed to be an independent cluster, with the number of repeated measurements $m_i$ ranging from 21 to 26.

Using the method of selecting the structural dimension in Supplementary Material \textcolor{blue}{S4}, the number of structural dimensions for the continuous predictors was chosen to be $d = 2$, while with only two binary predictors we set $d^\prime = 1$. Fitting the RMIR model in \eqref{eqn:modelqbinary_p1}--\eqref{eqn:modelqbinary_p3} produced the resulting estimate characterizing the overall fixed effects subspace on the continuous predictors
\begin{align*}
(\widehat{\bm\Delta}^{-1}\widehat{\bm\Gamma}_0)^\top = 
\begin{blockarray}{cccccc}
X_{ij1} & X_{ij2} & X_{ij3} & X_{ij4} & X_{ij5} & X_{ij6} \\ 
\begin{block}{[cccccc]}
-0.997 & 0.019 & 0.028 & 0.032 & 0.066 & -0.002 \\ 
  0.039 & -0.441 & 0.833 & 0.167 & 0.286 & -0.006 \\ 
\end{block}
\end{blockarray}.
\end{align*}
Across all countries, the first sufficient predictor for continuous predictors is influenced mostly by log income per capita, while the second sufficient predictor is heavily driven by sex ratio and infant mortality.
A basis for the dimension reduction space corresponding to the binary covariates is estimated to be $\hat{\bB} = (-0.06, 0.03, 0.99)^\top$, suggesting that the fixed effects central subspace for the sufficient predictor at the cluster level is dominated by the interaction $W_1W_2$. In Supplementary Material S6, we present plots of the response as a function of the estimated fixed effects sufficient predictors at the observation level, which suggests that across all the countries and years, the life expectancy of females exhibits a relatively strong but potentially non-linear relationship with at least one of the two sufficient predictors.

Next, the RMIR model produced the following estimate of the random effects covariance matrix for the continuous predictors
\begin{align*}
\widehat{\bm\Sigma} = 
\begin{blockarray}{ccc cccc}
& X_{ij1} & X_{ij2} & X_{ij3} & X_{ij4} & X_{ij5} & X_{ij6} \\ 
\begin{block}{c[cccccc]}
X_{ij1} & 0.001 & 0.001 & -0.001 & -0.000 & 0.001 & 0.001 \\ 
X_{ij2} &  0.001 & 0.005 & -0.005 & 0.000 & 0.009 & 0.004 \\ 
X_{ij3} & -0.001 & -0.005 & 0.006 & -0.000 & -0.010 & -0.007 \\ 
X_{ij4} & -0.000 & 0.000 & -0.000 & 0.000 & 0.001 & 0.001 \\ 
X_{ij5}  & 0.001 & 0.009 & -0.010 & 0.001 & 0.016 & 0.009 \\ 
X_{ij6} &  0.001 & 0.004 & -0.007 & 0.001 & 0.009 & 0.018 \\
\end{block}
\end{blockarray}.
\end{align*}
Examining its diagonal elements, we note children per woman and income inequality exhibited the largest variance estimates, suggesting they were responsible for driving heterogeneity among countries in terms of the cluster-specific random effect central subspaces. To further investigate this, we constructed cluster-specific random effects central subspaces $[\widehat{\bTheta}_i]$ given by RMIR, and summarized this by reporting the importance of each predictor for each subspace based on the corresponding diagonal elements of the estimated projection matrix. 
That is, we computed the $(p+q) \times (p+q)$ projection matrix $\mathcal{P}(\widehat{\bm\Theta}_i)$ and use its diagonal elements
to evaluate the importance of each covariate (averaged across both structural dimensions) for the $i$th country. Higher values of the diagonal elements in the projection matrix suggest the corresponding covariate is more important \citep[see][for examples of its usage elsewhere in SDR]{tan2018convex,nghiem2021sparse}.
Figure \ref{fig:centralsubspaces_importance} displays the resulting importance of each predictor for select countries.
Consistent with the estimated $(\widehat{\bm\Delta}^{-1}\widehat{\bm\Gamma}_0)^\top$ above, the cluster-specific central subspaces were strongly influenced by log income per capita, sex ratio, infant mortality, although interestingly some countries' central subspaces were also strongly driven by income inequality. The importance of the predictors varied greatly from one country to another: infant mortality was important in Egypt, but not in the Czech Republic and Laos, while the number of children per woman was quite important in Singapore but not in Romania, Bangladesh, etc. The two binary predictors were unimportant across most selected countries for driving the sufficient predictor at the observation level. 

\begin{figure}[t]
\centering
\includegraphics[width = \textwidth]{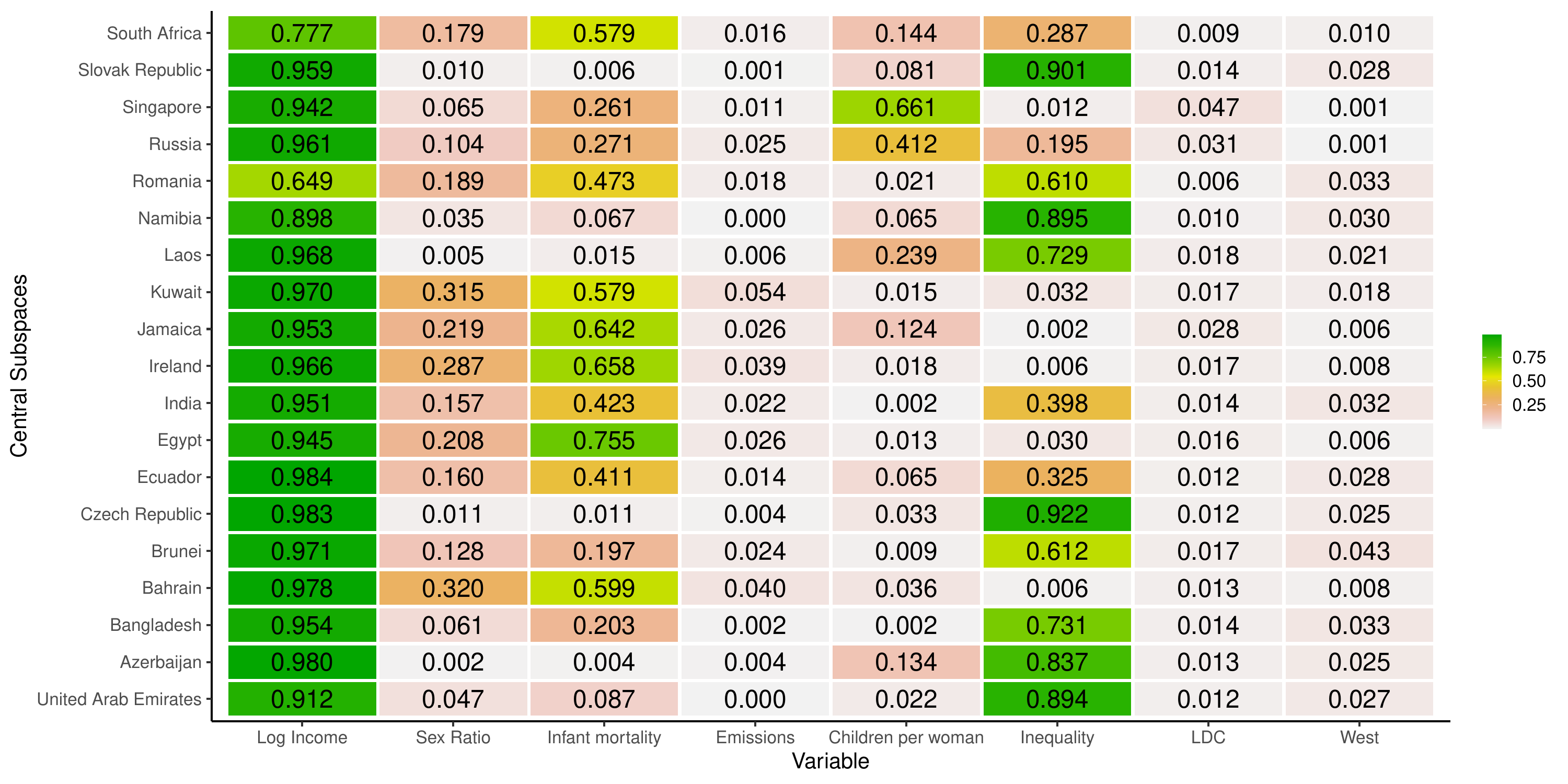}
\caption{Variable importance of the $p=6$ continuous predictors and $q=2$ time-invariant binary predictors in driving cluster-specific central subspaces at the observation level for select countries, as obtained from the RMIR model. An orange-green color gradient indicates variable importance, with dark green representing stronger importance.}
\label{fig:centralsubspaces_importance}
\end{figure}

Finally, in Supplementary Material S6 we present scatterplots of the female life expectancy versus the predicted cluster-specific sufficient predictors at the observation level (see below Proposition \ref{proposition:mixedcontinuous_qbinary} for this form) for select countries. Results show that female life expectancy exhibits a strong, sometimes close to linear relationship with one or both of these sufficient predictors across many countries, although they also verify the extent of heterogeneity between countries in terms of the random effects SDR operation.
}
\section{Discussion}
\label{sec:conclusion}
This article introduces the idea of random effects sufficient dimension reduction in independent clustered data settings, where the cluster-specific random effect central subspaces are assumed to follow a common distribution on a Grassmann manifold. 
\revise{
We incorporate random effects SDR first through a principal fitted components model for continuous predictors, with
asymptotic results and a numerical study demonstrating its strong performance against a global and separate fixed effects inverse regression models. We then extended the RPFC model to consider mixed predictors, focusing on random effects model-based inverse regression when we have continuous along with time-invariant or time-varying binary predictors, and establish the overall fixed effects and cluster-specific central subspaces in these scenarios. Finally, we performed random effects SDR to analyze the longitudinal relationship between female life expectancy and various socioeconomic variables across 117 countries, with results showing considerable heterogeneity across countries in the SDR process with only a small number of continuous predictors being important in this process. 
}

\revise{
This article makes an important first step towards generalizing the idea of SDR to longitudinal and independent clustered data. }
On the theoretical front, future research could investigate the prediction error of the cluster-specific central spaces e.g., the behavior of $n^{-1}\sum_{i=1}^n \Vert\mathcal{P}(\widehat{\bm\Theta}_i)- \mathcal{P}(\bm\Theta_i)\Vert_F$ when the cluster sizes $m_i$ are also growing $n$, 
and develop inferential properties for the parameters characterizing the fixed effects central subspace and the effects covariance matrix. \revise{Lower-level sufficient conditions can also be explored as part of establishing the large-sample properties of the proposed estimators. 
One can consider allowing different clusters to have different structural dimensions, although in this case since the cluster-specific central subspaces $[\bTheta_i]$ belong to Grassmann manifold $\Gr(p, d_i)$ with varying dimensions, it is more challenging to define a common random effects distribution. One approach to tackling this extension might be to incorporate regularization penalties that can also indirectly choose $d_i$ shrinking entire columns of $\mathbf{V}_i$ to zero in an adaptive, clusters-specific manner \cite[see][for example of such sparsity penalties for choosing structural dimension in the context of latent variable modeling]{hui2018order}. 
Another direction is to extend random effects SDR to moment-based inverse methods e.g, sliced inverse regression, and semiparametric estimating equations \citep{ma2012semiparametric}, though it is somewhat unclear how without an explicit objective function, heterogeneity defined on the tangent space can be straightforwardly incorporated into such SDR approaches.
Finally, the concept of random effects SDR and corresponding models developed in this paper can be potentially extended to multi-way clustered data \citep{mackinnon2021wild}, with a major challenge here being to address different clustering in a broad enough manner to cover crossed versus nested design. For example, with multilevel data, we speculate the random effects central subspaces themselves could be nested within each other, though it is unclear whether defining the nesting on the tangent subspaces suffices for this.}

\begin{center}
{\bf SUPPLEMENTARY MATERIALS}
\end{center}
The Supplementary Materials contains a review of Grassmann manifold, the proof of all theoretical results in Sections \ref{sec:randomeffectsCS}-\ref{sec:RMIR}, computational details for the RPFC and RMIR models in Sections \ref{sec:RPFC} and \ref{sec:RMIR} respectively, further simulation results for Section \ref{sec:simulation}, and additional application results for Section \ref{sec:dataapplication}. 

A Github repository with all the codes to implement the proposed estimation methods and instructions to reproduce the results throughout the paper are available at \url{https://github.com/lnghiemum/RMIR-paper}.  

\bibliographystyle{chicago} 
\bibliography{main}

\newpage

\appendix

\renewcommand{\thecondition}{SC\arabic{condition}}%

\renewcommand{\thefigure}{S\arabic{figure}}
\renewcommand{\thesection}{S\arabic{section}}

\begin{center}
\textbf{\Large Supplementary Materials for ``Random effects model-based sufficient dimension reduction for independent clustered data''}
\end{center}

\section{Background on Grassmann manifolds}
\label{section:review}

A Grassmann manifold $\Gr(p,d) $ is the set of all linear subspaces with dimension $d$ of $\mathbb{R}^p$. One way to represent a point on a Grassmann manifold is from a basis perspective \citep{bendokat2024grassmann}. That is, a subspace $\mathcal{U} \in \Gr({p, d})$ is identified by a non-unique, semi-orthogonal matrix $\bU \in \mathbb{R}^{p \times d}$, satisfying $\bU^\top \bU = \mathbf{I}_d$, whose columns form a basis for $\mathcal{U}$. To simplify the notation between a space and its semi-orthogonal matrix, we will write $\mathcal{U} = [\bU]$. 
A Grassmann manifold is often equipped with a Riemann metric, which roughly speaking is an inner product defined on the tangent space of the manifold. 
In more detail, given a subspace $[\bU] \in \Gr(p, d)$, the tangent space of $\Gr(p,d)$ at $[\bU]$, denoted as $T_{[\bU]} \Gr(p,d)$, is the collection of all possible directions 
of a curve on the manifold that passes through $[\bU]$. That is, $T_{[\bU]} \Gr(p,d) = \{\bV \in \mathbb{R}^{p \times d} \vert \bU^\top \bV = \mathbf{0} \}$ is the set of all matrices orthogonal to $\mathbf{U}$.
Equivalently, any matrix $\mathbf{V} \in T_{[\bU]} \Gr(p,d)$ can be written in the form $\mathbf{V} = (\mathbf{I}_p - \bU\bU^\top) \mathbf{A}$, where $\mathbf{A}$ is any arbitrary matrix in $\mathbb{R}^{p\times d}$. Importantly, this tangent space is a vector space: for any two matrices $\bV_1, \bV_2 \in T_{[\bU]} \Gr(p,d)$, we have $c_1 \bV_1 + c_2 \bV_2 \in T_{[\bU]} \Gr(p,d)$ for any two scalars $c_1$ and $c_2$. As such, we can embed the tangent space with an inner product, $\Phi(\bV_1, \bV_2) = 2^{-1} \text{trace}(\bV_1^\top \bV_2)$, which is known as a Riemann metric of the Grassmann manifold \citep{zimmermann2017matrix}. 

Next, with this Riemann metric we can define the distance between any two points $[\bU_1]$ and $[\bU_2]$ on the Grassmann manifold to be the length of the shortest curve between these two points (also known as the Riemann distance). Let $\alpha: [0,1] \rightarrow \Gr(p,d)$ be a curve on the Grassmann manifold, as parameterized by a scalar $t \in \mathbb{R}$, i.e $\alpha(0) = [\bU_1]$ and $\alpha(1) = [\bU_2]$.  As the curve passes through $[\bU_1]$, then its velocity vector $\dot{\alpha}(t) = d\alpha(t)/dt$ belongs to $T_{[\bU_1]} \Gr(p,d)$, and so the above Riemann metric can be used to define its length to be $L[\alpha]=\int_0^1 \Phi\left\{\dot{\alpha}(t), \dot{\alpha}(t)\right\} dt$. Thus, the distance between $[\bU_1]$ and $[\bU_2]$ is the infimum of this function $L$, and the curve which achieves this infimum is known as a geodesic. \citet{bendokat2024grassmann} gives the explicit formula for the Riemann distance between $[\bU_1]$ and $[\bU_2]$ to be $\text{dist}([\bU_1], [\bU_2]) = \left(\sum_{i=1}^{d} \sigma_d^2\right)^{1/2}$, where $\sigma_i$ is the $i$th largest singular value of the $d\times d$ matrix $\bU_1^\top \bU_2$ for $i=1,\ldots, d$. As a result, the upper bound for this distance is $\text{dist}([\bU_1], [\bU_2]) \leq \sqrt{d}\pi/2$.   

Since the tangent space is a vector space, then we will define a probability distribution in this space whose variability subsequently determines the variability of points on the manifold. Furthermore, we will define a mapping between points on the tangent space and points on the Grassmann manifold. For the latter, we employ the exponential mapping defined as follows: given a point $[\bU_1] \in \Gr(p,d)$ and a point $\bV \in T_{[\bU_1]} \Gr(p,d)$, the exponential map $\operatorname{Exp}_{[\bU_1]}$ transfer a point $\bV \in T_{[\bU_1]} \Gr(p,d)$  to $[\bU_2] \in \Gr(p,d)$. As shown in \citet{bendokat2024grassmann}, a semi-orthogonal basis for $[\bU_2]$ is explicitly given by
$\bU_2 = \operatorname{Exp}_{[\bU_1]}(\bV) = \bU_1\mathbf{M} + \mathbf{Q}\mathbf{N}$, where $\mathbf{QR}= \bV$ is the QR decomposition of $\bV$, $\mathbf{M}  = \mathbf{D}\cos(\bm\Theta) \mathbf{D}^\top$, and $\mathbf{N} = \bm{\Phi}\sin(\bm\Theta) \mathbf{D}^\top$ where $\bm\Phi \bm\Theta\mathbf{D}^\top$ is the singular value decomposition (SVD) for $\mathbf{R}$. Note since $\mathbf{V}$ is orthogonal to $\mathbf{U}_1$, then so is $\mathbf{Q}$. Conversely, the inverse exponential mapping $\operatorname{Exp}^{-1}$ transfers a point $[\bU_2] \in \Gr(p, d)$ to a point $\bV \in T_{[\bU_1]}\Gr(p,d)$. The explicit formula for this is $\bV = \operatorname{Exp}^{-1}([\bU_2]) = \mathbf{Q}^*\arctan(\bm\Sigma^*)\mathbf{D}^\top$
where $\mathbf{Q}^*\bm\Sigma^*\mathbf{D}^\top$ is the SVD of $(\mathbf{I}_p - \bU\bU_1)\bU_2(\bU_1^\top \bU_2)^{-1}$.
In these above formulas, the cosine, sine, and inverse tangent functions are applied point-wise to the diagonal elements of corresponding matrices. 

\revise{Note the exponential mapping does not cover the entire Grassmann manifold. \citet{zimmermann2017matrix} proved that for any image $[\bU_2]$ of the exponential map from the tangent space $[\bU_1]$, the Riemann distance (defined in the previous paragraph) $\text{dist}([\bU_1], [\bU_2]) \leq \pi/2$; in this sense, the exponential mapping enables us to stay on a local neighborhood of $[\bGamma_0]$.}   

\revise{Finally, given a sample $\bGamma_1,\ldots,\bGamma_n \in \Gr(p,d)$, the sample Fr\'{e}chet mean  $[\widehat{\bGamma}_0^*]$ is defined to be
$$
[\widehat{\bGamma}_0^*] = \argmin_{[\bGamma_0] \in \Gr(p,d)} \dfrac{1}{n} \sum_{i=1}^{n} \text{dist}\left([\bGamma_i], [\bGamma_0] \right)
$$
and the sample Fr\'{e}chet variance is $n^{-1} \sum_{i=1}^{n} \text{dist}([\bGamma_i], [\widehat{\bGamma}_0])$. Note that this Fr\'{e}chet variance is upper bounded by $\sqrt{d}\pi/2$. Both Fr\'{e}chet mean and variance on the Grassmann manifold can be computed using the \texttt{R} package \texttt{manifold}  \citep{manifold2022package}. 
}

\section{Proof and derivations}
\subsection{Proof of Lemma 2.1}
To simplify the notation, we omit $i$ as a subscript in this proof. By the property of exponential mapping, we can write $\bm\Gamma = \bm\Gamma_0 \mathbf{M} + \mathbf{Q} \mathbf{N}$, where $\mathbf{QR} = \bV$ is the QR decomposition of $\bV$, $\mathbf{M}  = \mathbf{D}\cos(\bm\Sigma) \mathbf{D}^\top$, and $\mathbf{N} = \bm{\Psi}\sin(\bm\Sigma) \mathbf{D}^\top$, where
$\bm\Psi \bm\Sigma \mathbf{D}^\top$ is the singular matrix decomposition for $\mathbf{R}$. Note $\mathbf{M}$ and $\mathbf{N}$ are $d\times d$ matrices, and $[\bm\Gamma_0 \mathbf{M}] = [\bm\Gamma_0]$ for any $\mathbf{M}$. It can then be seen from the equation $\bm\Gamma_i = \Exp_{[\bm\Gamma_0]}(\bV_i), \bV_i \in T_{[\bm\Gamma_0]}\Gr(p, d)$ that $\mathbf{Q}$ is an even function and $\mathbf{N}$ is an odd function of $\mathbf{V}$. Therefore, $\mathbf{QN}$ is an odd function of $\mathbf{V}$, and since $\mathbf{V}$ has a symmetric density around zero, then we have $E(\mathbf{QN}) = \mathbf{0}$.  Hence $E(\bm\Gamma) = \bm\Gamma_0 E(\mathbf{M})$, which implies $[E(\bm\Gamma)] = \bm\Gamma_0$ as required.

\revise{
\subsection{Proof of Proposition 3.1}

First we prove that $[\bm\Theta_i] = \bm\Delta^{-1}\bGamma_i$ is the cluster-specific central subspaces for the $i$th cluster. It  suffices to prove that $y_{ij} \perp \bX_{ij} \mid (\bGamma_i^\T \bm\Delta^{-1}\bX_{ij}, \bGamma_i)$, i.e the conditional distribution of $\bX_{ij} \mid (y_{ij}, \bGamma_i^\T \bm\Delta^{-1}\bX_{ij}, \bGamma_i) $ does not depend upon $y_{ij}$. 

From equation  (4a) in the paper, we have $\bX_{ij} \mid (y_{ij}, \bGamma_i) \sim N_p(\bmu_i + \bGamma_i \bv_{ijy}, \bm\Delta)$, hence 
\begin{equation*}
\begin{pmatrix}
\bX_{ij} \\
\bGamma_i^\T \bm\Delta^{-1}\bX_{ij}
\end{pmatrix} ~ \Bigg| ~ (y_{ij}, \bGamma_i ) \sim N_{p+d}\left(\begin{pmatrix} \bmu_i + \bGamma_i \bv_{ijy} \\ 
\bGamma_i^\top\bm\Delta^{-1} \bmu_i + \bGamma_i^\top\bm\Delta^{-1}\bGamma_i \bv_{ijy}
\end{pmatrix}, \begin{pmatrix} \bm\Delta & \bGamma_i \\ \bGamma_i^\top & \bGamma_i^\top\bm\Delta^{-1} \bGamma_{i} \end{pmatrix}\right)
\end{equation*}
Using the property of the multivariate normal distribution, we then have  $\bX_{ij} \mid (y_{ij}, \bGamma_i^\T \bm\Delta^{-1}\bX_{ij}, \bGamma_i)$ follows a multivariate normal distribution with conditional mean and variance
\begin{equation}
\small
\begin{alignedat}{2}
& E\left\{\bX_{ij} \mid (y_{ij}, \bGamma_i^\T \bm\Delta^{-1}\bX_{ij}, \bGamma_i) \right\} && = \bmu_i + \bGamma_i \bv_{ijy} + \bGamma_i \left(\bGamma_i^\top\bm\Delta^{-1} \bGamma_{i}\right)^{-1} \left( \bGamma_i^\T \bm\Delta^{-1}\bX_{ij} -  \bGamma_i^\top\bm\Delta^{-1} \bmu_i- \bGamma_i^\top\bm\Delta^{-1}\bGamma_i \bv_{ijy} \right), 
\\
&   && = \bmu_i + \bGamma_i \left(\bGamma_i^\top\bm\Delta^{-1} \bGamma_{i}\right)^{-1}  \bGamma_i^\T \bm\Delta^{-1}(\bX_{ij}-\bmu_i) \\
& \Var\left\{\bX_{ij} \mid (y_{ij}, \bGamma_i^\T \bm\Delta^{-1}\bX_{ij}, \bGamma_i) \right\} && = \bm\Delta - \bGamma_i \left(\bGamma_i^\top\bm\Delta^{-1} \bGamma_i \right)\bGamma_i^\top,
\end{alignedat}
\end{equation}
respectively. Both the above conditional mean and variance do not depend on $y_{ij}$, so the conditional distribution $\bX_{ij} \mid (y_{ij}, \bGamma_i^\T \bm\Delta^{-1}\bX_{ij}, \bGamma_i) $ does not depend upon $y_{ij}$. Hence, $[\bTheta_i] = \bm\Delta^{-1}[\bGamma_i]$ is the $i$th cluster-specific central subspace. Combining this result with Lemma 2.1, we have $[\bTheta_0] = \bm\Delta^{-1}[\bGamma_0]$ is the overall mean central subspace. 
}
\subsection{Proof of Proposition 3.2}

It suffices to prove that $\ell_i(\bm\Gamma_0, \bm\mu_i, \bC, \bm\Sigma) = \ell_i(\bm\Gamma_0 \bA, \bm\mu_i, \bA^\top\bC, \bm\Sigma)$. We have
\begin{equation}
\ell(\bm\Gamma_0 \bA, \bm\mu_i, ~\bA^\top\bC, \bm\Sigma) =  \int \left\{\prod_{j=1}^{m_i} \text{N}(\bX_{ij}; \bm\mu_i + h(\bm\Gamma_0 \bA, \bV_i) \bA^\top \bC \boldf_{ijy}, \bm\Delta)\right\} \text{MN}(\bV_i; \mathbf{0}, \bm\Gamma, \bI_d) d\bV_i, 
\label{eq:integral}
\end{equation}

Let $\bU_i = \bV_i \bA$, or $\bV_i = \bU_i \bA^\top$. Since $\bV_i \in T_{[\bm\Gamma_0]}\Gr(d, p)$, then so is $\bU_i$. As a result, the limit of the integration on the right-hand side of \eqref{eq:integral} is unchanged when we express the integration in terms of $\bU_i$. Direct calculation verifies that $h(\bm\Gamma_0 \bA, \bV_i) = h(\bm\Gamma_0 \bA, \bU_i \bA^\top) = h(\bm\Gamma_0 , \bU_i) \bA$  . Furthermore, from the the joint density of $\bV_i$ given in the main text, we have $\text{MN}(\bU_i \bA^\top; \mathbf{0}, \bm\Gamma, \bI_d) = \text{MN}(\bU_i; \mathbf{0}, \bm\Gamma, \bI_d)$. Hence by the substitution rule, we have
$$
\begin{aligned}
\ell(\bm\Gamma_0 \bA, \bm\mu_i, ~\bA^\top\bC, \bm\Sigma) & =  \int \left\{\prod_{j=1}^{m_i} \text{N}(\bX_{ij}; \bm\mu_i +  h(\bm\Gamma_0 , \bU_i) \bA \bA^\top \bC \boldf_{ijy}, \bm\Delta)\right\} \text{MN}(\bU_i; \mathbf{0}, \bm\Gamma, \bI_d) ~ \vert \bA \vert ~ d\bU_i \\  &   = \int \left\{\prod_{j=1}^{m_i} \text{N}(\bX_{ij}; \bm\mu_i +  h(\bm\Gamma_0 , \bU_i) \bC \boldf_{ijy}, \bm\Delta)\right\} \text{MN}(\bU_i; \mathbf{0}, \bm\Gamma, \bI_d) d\bU_i \\
& \stackrel{(i)}{=} \ell(\bm\Gamma_0, \bm\mu_i, \bC, \bm\Sigma).
\end{aligned}
$$
where step $(i)$ follows from the property that $\vert \bA \vert = 1$ and $\bA\bA^\top = \bI_d$ for any orthogonal matrix $\bA$. The required result follows from this.

\subsection{Proof of Theorem 4.1}
In the proof below, we use the superscript `$\GPFC$' to denote estimators of the parameters in the global PFC model based on maximizing equation (6) in the main text. Furthermore, as it often the case with inverse regression methods, we treat $y_{ij}$ as non-stochastic. Let $\mathcal{X}$ denote the $N \times p$ matrix formed by stacking the $\bX_{ij}$'s as row vectors, and similarly define $\mathcal{F}$ as the $N \times r$ matrix formed by stacking the $\bm{f}_{ijy}$'s as row vectors, where the stacking is ordered by observations within clusters. 

The remainder of the proof hinges on the fact that, by construction of the global PFC model, the span of the matrix estimator $[\hat{\bC}^{\GPFC}]$ is equivalent to the span of the best rank-$d$ estimator for the coefficient matrix of a multivariate reduced rank regression where $\mathcal{X}$ is the full response matrix, $\mathcal{F}$ is the full covariate matrix, and $\tilde{\bm\Delta}$ is the assumed residual covariance matrix for each $i=1,\ldots, n; j=1,\ldots, m_i$. 

Using results for multivariate reduced rank regression then \citep[Chapter 2, p.34][]{reinsel2022multivariate}, we obtain
$
[\widehat{\bC}^{\GPFC}] = [\widehat{\bm\Sigma}_{xx}^{1/2}\hat{\bm\Lambda}],
$
where 
\begin{itemize}[leftmargin=*]
    \item $\widehat{\bm\Sigma}_{xx} = N^{-1} \sum_{i=1}^{n}\sum_{j=1}^{m_i}\bX_{ij}\bX_{ij}^\top = N^{-1}\mathcal{X}^\top\mathcal{X}$;
    \item $\widehat{\bm\Sigma}_{ff} = N^{-1} \sum_{i=1}^{n}\sum_{j=1}^{m_i}\boldf_{ijy}\boldf_{ijy}^\top = N^{-1} \mathcal{F}^\top\mathcal{F}$;
    \item $\widehat{\bm\Sigma}_{xf} = \widehat{\bm\Sigma}_{fx}^\top = N^{-1}\sum_{i=1}^{n}\sum_{j=1}^{m_i}(\bX_{ij} - \overline{\bX})\boldf_{ijy}^\top$, with $
    \overline{\bX} = N^{-1} \sum_{i=1}^{n}\sum_{j=1}^{n_i}\bX_{ij}   $;
    \item $\hat{\bm\Lambda}$ is the $p \times d$ matrix formed from the first $d$ normalized eigenvectors of $\widehat{\bm\Sigma}_{xx}^{-1/2}\hat{\bm\Sigma}_{xf} \widehat{\bm{\Sigma}}_{ff}^{-1} \widehat{\bm\Sigma}_{fx} \widehat{\bm\Sigma}_{xx}^{-1/2}$.
     \end{itemize}

Since $N \to \infty$ as $n \rightarrow \infty$ (assuming the cluster sizes $m_i$ are bounded), then by the law of large numbers we have
\begin{equation}
\widehat{\bm\Sigma}_{xf} = \dfrac{1}{N} \sum_{i=1}^{n}\sum_{j=1}^{m_i} E\left\{(\bX_{ij} - \overline{\bX}_i)\boldf_{ijy}^\top\right\} + O_p\left(N^{-1/2} \right).
\label{eq:lln}
\end{equation}
Next, by the law of iterated expectation, we obtain
$$
\begin{aligned}
E\left\{(\bX_{ij} - \overline{\bX}_i)\boldf_{ijy}^\top\right\} & = E\left[ E\left\{(\bX_{ij} - \overline{\bX}_i)\boldf_{ijy}^\top \mid \bGamma_i \right\} \right] \\
& =  E \left( \bGamma_i   \right) \bC\boldf_{ijy} \boldf_{ijy}^\top \\
& \stackrel{(i)}{=} \bGamma_0 \mathbf{M} \bC\boldf_{ijy} \boldf_{ijy}^\top,
\end{aligned}
$$
where step $(i)$ follows from the argument in the derivation of Lemma 1 with $\mathbf{M}$ as a (deterministic) non-singular $d\times d$ matrix. Substituting this result into equation~\eqref{eq:lln}, we obtain
$$
\widehat{\bm\Sigma}_{xf} = \bGamma_0 \mathbf{M} \bC \left(\dfrac{1}{N} \sum_{i=1}^{n}\sum_{j=1}^{m_i} \boldf_{ijy} \boldf_{ijy}^{\top}  \right) +O_p\left(N^{-1/2} \right) = \bGamma_0 \mathbf{M} \bC \widehat{\bm\Sigma}_{ff} +O_p\left(N^{-1/2} \right).
$$
As $n \to \infty$ then, then under Condition C1 we obtain
$$
\widehat{\bm\Sigma}_{xf}\widehat{\bm\Sigma}_{ff}^{-1} \widehat{\bm\Sigma}_{fx} = \bGamma_0 \mathbf{M} \bC \bm\Sigma_{ff} \bC^{\top}\mathbf{M}^\top \bGamma_0^\top +O_p\left(N^{-1/2} \right).
$$
Finally, we note $\bGamma_0 \mathbf{M} \bC \bm\Sigma_{ff} \bC^{\top}\mathbf{M}^\top \bGamma_0^\top$ is a $p \times p $ matrix of rank $d$, as is $\bm\Sigma_{xx}^{-1/2} \bGamma_0 \mathbf{M} \bC \bm\Sigma_{ff} \bC^{\top}\mathbf{M}^\top \bGamma_0^\top \bm\Sigma_{xx}^{-1/2}$ Hence, as $n \to \infty$ and under Condition C2, the space spanned by its $d$ eigenvectors is equivalent to its column space. It follows that 
$$
[\widehat{\bC}^{\GPFC}] \stackrel{p}{\to} \bm\Sigma_{xx}^{1/2}[\bm\Sigma_{xx}^{-1/2} \bGamma_0 \mathbf{M} \bC \bm\Sigma_{ff} \bC^{\top}\mathbf{M}^\top \bGamma_0^\top \bm\Sigma_{xx}^{-1/2}] = [\bm\Gamma_0],
$$
as required.

\subsection{Proof of Theorem 4.2}
Recall the estimated $\widehat{\bm\Psi} =  \{\widehat{\bC}^\top, \text{vech}(\widehat{\bm\Delta})^\top, \text{vech}(\widehat{\bm\Sigma})^\top\}^\top$ from the MCEM algorithm maximizes the marginal log-likelihood 
$\ell(\bm{\Psi} \mid {\bGamma_0^* \bA}) = n^{-1} \sum_{i=1}^{n} \log L_{i}(\bm\Psi \mid {\bGamma_0^* \bA})$, 
where $L_{i}(\bm\Psi \mid {\bGamma_0^* \bA}) = f(\bZ_i \mid {\bGamma_0^* \bA}, \bm\Psi) =   \int \text{MN}(\bZ_{i};\bm\Gamma_i \bC\bH_{iy}, \bm\Delta, \bL_i) \, \text{MN}(\mathbf{V}_i; \mathbf{0}, \bm\Sigma, \mathbf{I}_d)  d\mathbf{V}_i$. By the law of large numbers for independent but not identically distributed random variables,  as $n \to \infty$, for any $\bm\Psi$, we have
$\ell(\bm\Psi \mid {\bGamma_0^* \bA}) =\ell_*(\bm\Psi \mid \bm\Gamma_0^*\bA) + O_p(n^{-1/2})$, where $\ell_{*}(\bm\Psi \mid \bGamma_0^*\bA) = E_{\tilde{\bm\Psi}^*}\left[\ell(\bm{\Psi} \mid \bGamma_0^* \bA) \right]$ and the expectation is with respect to the true value $\widetilde{\bm\Psi}^*$ of the data $\bZ_1, \ldots, \bZ_n$. We will prove that $\ell_*(\bm\Psi \mid \bGamma_0^*\bA)$ is maximized at $\bm\Psi^* = (\bA^\top\bC^*, \bm\Delta^*, \bm\Sigma^*)$, i.e. $\ell_*(\bPsi \mid \bGamma_0^*\bA) - \ell_*(\bPsi^* \mid \bGamma_0^*\bA) \leq 0$ uniformly for all $\bm\Psi$. Indeed,  
$$
\begin{aligned}
\ell_*(\bPsi \mid \bGamma_0^*\bA) - \ell_*(\bPsi^* \mid \bGamma_0^*\bA) & = E_{\tilde{\bm\Psi}^*}\left[\ell(\bm{\Psi} \mid \bGamma_0^* \bA) - \ell(\bm{\Psi}^* \mid \bGamma_0^* \bA)  \right] \\
& = \dfrac{1}{n} \sum_{i=1}^{n} E_{\tilde{\bm\Psi}^*} \left[ \log \dfrac{L_i(\bm\Psi \mid \bm\Gamma_0^*\bA)}{L_i(\bm\Psi^* \mid \bm\Gamma_0^*\bA)} \right] \\
& \stackrel{(i)}{\leq} \left\{ \dfrac{1}{n}  \sum_{i=1}^{n} E_{\tilde{\bm\Psi}^*}  \left[ \dfrac{L_i(\bm\Psi \mid \bm\Gamma_0^*\bA)}{L_i(\bm\Psi^* \mid \bm\Gamma_0^*\bA)}  \right] \right\} - 1 \\
& = \dfrac{1}{n}\sum_{i=1}^{n} \left\{ \int \dfrac{f(\bZ_i \mid  \bGamma_0^*\bA, \bPsi)}{f(\bZ_i \mid  \bGamma_0^*\bA, \bPsi^*)} f(\bZ_i \mid  \widetilde{\bPsi}^*) d\bZ_i \right\} - 1
\end{aligned},
$$
where step $(i)$ follows from the Jensen's inequality.
By a similar argument to the proof of Proposition 1, we have $f(\bZ_i \mid  \bGamma_0^*\bA, \bPsi^*) = f(\bZ_i \mid  \widetilde{\bPsi}^*)$. Therefore,
$$
\ell_*(\bPsi \mid \bGamma_0^*\bA) - \ell_*(\bPsi^* \mid \bGamma_0^*\bA)  \leq \dfrac{1}{n}\sum_{i=1}^{n} \left\{ \int f(\bZ_i \mid  \bGamma_0^*\bA, \bPsi) d\bZ_i \right\} -1 = 0,
$$
as required. Hence, when the regularity conditions are satisfied, consistency of $\hat{\bm\Delta}$ and $\widehat{\bm\Sigma}$ follow from the same argument as in  \citet[Section 6.5]{lehmann2006theory}.

\revise{
\subsection{Proof of Proposition 6.1}
In this proof, we use the same notation as \citet{bura2023sufficient}. Particularly,  for any $m \times n$ matrix $\mathbf{G}$, the vectorization $\text{vech}(\mathbf{G})$ is the $m n \times 1$ vector obtained by stacking the columns of $\mathbf{G}$. The half-vectorization operator (vech) converts the lower half of a square matrix including the main diagonal to a vector. That is, if $\mathbf{G}$ is a square $q \times q$ matrix then $\operatorname{vech}(\mathbf{G})$ is a $q(q+1)/2 \times 1$ vector obtained by stacking the columns of the lower triangular part of $\mathbf{G}$ including the diagonal. There is a unique $\mathbf{D}_q \in \mathbb{R}^{q^2 \times q(q+1) / 2}$ and $\mathbf{C}_q \in \mathbb{R}^{q(q+1) / 2 \times q^2}$, such that $\operatorname{vec}(\mathbf{G})=\mathbf{D}_q \operatorname{vech}(\mathbf{G})$ and $\operatorname{vech}(\mathbf{G})=\mathbf{C}_q \operatorname{vec}(\mathbf{G})$. We let the matrix $\mathbf{L}_q \in \mathbb{R}^{q \times q(q+1) / 2}$ has entries $1$ and $0$, so that $\mathbf{L}_q \mathbf{C}_q$ is equal to $\mathbf{C}_q$ except for replacing the value $1 / 2$ by zero throughout. The matrix $\mathbf{J}_q \in \mathbb{R}^{k_q \times q(q+1) / 2}$ has entries 1 and 0 , so that $\mathbf{J}_q\mathbf{C}_q$ is equal to $\mathbf{C}_q$ except for replacing the ones with zeros. 

The key step in this proof is to write the joint distribution $(\bX_{ij}, \bW_i)$ in the quadratic exponential family form as in \citet{bura2015sufficient} and \citet{bura2023sufficient}. To this end, we have
$$
\begin{aligned}
& p(\bX_i, \bW_i \mid \bm{y}_i, \bGamma_i) = \left\{\prod_{j=1}^{m_i} p(\bX_{ij} \mid (y_{ij}, \bW_{i}, \bGamma_i) \right\} p(\bW_i \mid \bm{y}_i) \\
& \propto \exp\left\{-\dfrac{1}{2} \sum_{j=1}^{m_i} \left(\bX_{ij} - \bm\mu - \bGamma_i\bC_1 \boldf_{ijy} - \bm\beta \bW_i  + \bm\beta \bmu_W\right)^\T \bm\Delta^{-1} \left(\bX_{ij} - \bm\mu - \bGamma_i\bC_1 \boldf_{ijy} - \bm\beta \bW_i + \bm\beta \bmu_W \right)  \right\} \\
& \times \exp\left\{\operatorname{vech}^\T(\bW_i\bW_i^\T) \operatorname{vech}^\T\left(\bm\tau_0 + \bA \bC_2 \bm{g}_{i\bm{y}}\right)\right\} \\
\end{aligned}
$$
We next match it to the quadratic exponential family form, i.e $$ p(\bX_i, \bW_i \mid \bm{y}_i, \bGamma_i) \propto  \exp\left\{\bT^\top(\bX_{i}, \bW_i)\boldeta_{i\bm{y}} - \psi\left(\boldeta_{yi}\right)\right\}, $$ where $\bT(\bX_{i}, \bW_i)$ is an appropriate sufficient statistic, $\boldeta_{i\bm{y}}$ is a natural parameter, and $\psi\left(\boldeta_{i\bm{y}}\right)$ is a conditional cumulant function.  To do that, we re-write 
\begin{align*}
\begin{aligned}
& -\dfrac{1}{2} \sum_{j=1}^{m_i} \left(\bX_{ij} - \bm\mu - \bGamma_i\bC_1 \boldf_{ijy} - \bm\beta \bW_i  + \bm\beta \bmu_W \right)^\T \bm\Delta^{-1} \left(\bX_{ij} - \bm\mu - \bGamma_i\bC_1 \boldf_{ijy} - \bm\beta \bW_i  + \bm\beta \bmu_W \right) \\ & +\operatorname{vech}^\T(\bW_i\bW_i^\T) \operatorname{vech}^\T\left(\bm\tau_0 + \bA \bC_2 \bm{g}_{i\bm{y}}\right) \\ = &  
 -\dfrac{1}{2} \sum_{j=1}^{m_i}\bX_{ij}^\T \bm\Delta^{-1}\bX_{ij} + \sum_{j=1}^{m_i} \bX_{ij}^\T \bm\Delta^{-1} \bm\mu + \sum_{j=1}^{m_i} \bX_{ij}^\T \bm\Delta^{-1} \bGamma_i \bC_1\boldf_{ijy} + \sum_{j=1}^{m_i} \bX_{ij}^\T \bm\Delta^{-1} \bm\beta \bW_{i}  -\sum_{j=1}^{m_i} \bX_{ij}^\top \bm\Delta^{-1}\bm\beta \bmu_W  \\
& - \dfrac{m_i}{2} \bm\mu^\T \bm\Delta^{-1}\bm\mu -  \bm\mu^\T \bm\Delta^{-1} \bm\Gamma_i \bC_1 \sum_{j=1}^{m_i}\boldf_{ijy} - m_i\bm\mu^\T \bm\Delta^{-1}\bm\beta\bW_i + m_i \bmu\bm\Delta^{-1}\bm\beta\bmu_W \\
& - \dfrac{1}{2}\sum_{j=1}^{m_i} \boldf_{ijy}^\T \bC_1^\top \bGamma_i^\T \bm\Delta^{-1}\bGamma_i \bC_1 \boldf_{ijy}   -  \sum_{j=1}^{m_i} \boldf_{ijy}^\T \bC_1^\T\bm\Gamma_i^\T\bm\Delta^{-1} \bm\beta\bW_i +  \sum_{j=1}^{m_i} \boldf_{ijy}^\T \bC_1^\T\bm\Gamma_i^\T\bm\Delta^{-1} \bm\beta\bmu_W \\ &
- \dfrac{m_i}{2} \bW_i^\T \bm\beta^\T \bm\Delta^{-1} \bm\beta \bW_i + m_i \bW_i^\T \bm\beta^\T \bm\Delta^{-1} \bm\beta \bmu_W -\dfrac{m_i}{2}\bmu_W^\top \bm\beta^\top \bm\Delta^{-1}\bm\beta\bmu_W  \\&+\operatorname{vech}^\T(\bW_i\bW_i^\T) \bm\tau_0 + \operatorname{vech}^\T(\bW_i\bW_i^\T) \bA\bC_2 \bm{g}_{i\bm{y}}.
\end{aligned}]
\end{align*}
Next, following the same arrangement as in \citet{bura2023sufficient} and let $\bF_{yi}$ be the $m_i \times r_1$ matrix whose $j$th row is $\bm{f}_{ijy}$, we have

\begin{equation*}
\begin{aligned}
& \mathbf{T}^T(\bX_i, \bW_i) \boldsymbol{\eta}_{i\bm{y}} \\ & =\sum_{j=1}^{m_i}\bX_{ij}^T  \boldsymbol{\Delta}^{-1} \boldsymbol{\mu}+\sum_{j=1}^{m_i}\bX_{ij}^T \boldsymbol{\Delta}^{-1} \bGamma_i \bC_1 \boldf_{ijy}  -\sum_{j=1}^{m_i} \bX_{ij}^\top \bm\Delta^{-1}\bm\beta \bmu_W  \\ & -m_i\bW_i^\T \boldsymbol{\beta}^T \boldsymbol{\Delta}^{-1} \boldsymbol{\mu}+\bW_i^T \boldsymbol{\beta}^T \boldsymbol{\Delta}^{-1} \bGamma_i\bC_1 \sum_{j=1}^{m_i}\boldf_{ijy} + m_i \bW_i^\T \bm\beta^\T \bm\Delta^{-1} \bm\beta \bmu_W  -\frac{1}{2} \sum_{j=1}^{m_i}\bX_{ij}^\T \boldsymbol{\Delta}^{-1} \bX_{ij} \\ & + \sum_{j=1}^{m_i}\bX_{ij}^\T\boldsymbol{\Delta}^{-1} \boldsymbol{\beta} \bW_i -\frac{m_i}{2} \bW_i^T \boldsymbol{\beta}^T \boldsymbol{\Delta}^{-1} \boldsymbol{\beta} \bW_i  +\operatorname{vech}^T\left(\bW_i \bW_i^\T\right) \boldsymbol{\tau}_0+\operatorname{vech}^T\left(\bW_i \bW_i^\T\right) \bA\bC_2 \bm{g}_{i\bm{y}} \\
& = \left\{\operatorname{vec}\left(\bX_i^\T\right) \right\}^\T \left\{\operatorname{vec}\left(\bm\Delta^{-1}\bm\mu \bm{1}_{m_i}^\T \right) - \operatorname{vec}\left(\bm\Delta^{-1}\bm\beta
\bmu_W\bm{1}_{m_i}^\T \right) + \left(\mathbf{F}_{yi} \otimes \bI_p \right)\operatorname{vec}\left(\bm\Delta^{-1}\bGamma_i \bC_1 \right) \right\} 
\\ & + \bW_i^\T \left\{-m_i \bm\beta^\T \bm\Delta^{-1}\bm\mu + m_i \bm\beta^\T \bm\Delta^{-1}\bm\beta\bm\mu_W + \left( \sum_{j=1}^{m_i} \boldf_{ijy} \right) \operatorname{vec}\left(\bm\beta^\T\bm\Delta^{-1}\bGamma_i \bC_1 \right) \right\} \\
& -\frac{1}{2}\left\{\mathbf{D}_p \mathbf{D}_p^T \operatorname{vech}\left(\sum_{j=1}^{m_i} \bX_{ij}\bX_{ij}^\T\right)\right\}^\T \operatorname{vech}\left(\boldsymbol{\Delta}^{-1}\right)+\operatorname{vec}\left(\sum_{j=1}^{m_i}\bX_{ij}^\T \bW_i\right)^\T \operatorname{vec}\left(\boldsymbol{\Delta}^{-1} \boldsymbol{\beta}\right) \\
& +\bW_i^T\left\{-\frac{m_i}{2} \mathbf{L}_q \mathbf{D}_q^T \operatorname{vec}\left(\boldsymbol{\beta}^T \boldsymbol{\Delta}^{-1} \boldsymbol{\beta}\right)+\mathbf{L}_q \boldsymbol{\tau}_0
\right\} \\
& +\left\{\mathbf{J}_q \operatorname{vech}\left(\bW_i\bW_i^\T\right)\right\}^T\left\{-\frac{1}{2} \mathbf{J}_q \mathbf{D}_q^T \operatorname{vec}\left(\boldsymbol{\beta}^\T \boldsymbol{\Delta}^{-1} \boldsymbol{\beta}\right)+\mathbf{J}_q \boldsymbol{\tau}_0+\left(\bm{g}_{i\bm{y}}^\T \otimes \mathbf{I}_{k_q}\right) \operatorname{vec}\left(\mathbf{J}_q \bA\bC_2\right)\right\} 
\\ & =\left\{\operatorname{vec}\left(\bX_i^\T\right) \right\}^\T  \boldsymbol{\eta}_{i\bm{y} 1}+\bW_i^T \boldsymbol{\eta}_{i\bm{y}2}-\frac{1}{2}\left(\mathbf{D}_p^T \mathbf{D}_p \operatorname{vec}\left(\bX_i^\T \bX_i\right)\right)^T \boldsymbol{\eta}_3 +\operatorname{vec}\left(\sum_{j=1}^{m_i}\bX_{ij}\bW_i^\T\right)^\T \boldsymbol{\eta}_4 \\ & + \left(\mathbf{J}_q \operatorname{vech}\left(\bW_i \bW_i^\T\right)\right)^\T \boldsymbol{\eta}_{i\bm{y} 5},
\end{aligned}
\end{equation*}
with $\bm\eta_{i\bm{y}} = (\boldeta_{i\bm{y}1}^\T, \boldeta_{i\bm{y}2}^\T, \boldeta_{3}^\T,\boldeta_{4}^\T,\boldeta_{i\bm{y}5}^\T)^\T $ where
$$
\begin{aligned}
\bm\eta_{i\bm{y}1} & = \operatorname{vec}\left(\bm\Delta^{-1}\bm\mu \bm{1}_{m_i}^\T \right) - \operatorname{vec}\left(\bm\Delta^{-1} \bm\beta\bmu_W \bm{1}_{m_i}^\T \right)  + \left(\mathbf{F}_{yi} \otimes \bI_p \right)\operatorname{vec}\left(\bm\Delta^{-1}\bGamma_i \bC_1 \right)  \\
\boldsymbol{\eta}_{i\bm{y}2} &= m_i \boldsymbol{\beta}^T \boldsymbol{\Delta}^{-1} \boldsymbol{\mu} + m_i \bm\beta^\top \bm\Delta^{-1}\bm\beta\bmu_W +  \sum_{j=1}^{m_i} \boldf_{ijy}\operatorname{vec}\left(\boldsymbol{\beta}^T \boldsymbol{\Delta}^{-1} \bGamma_i \bC_1\right) \\
& -\frac{m_i}{2} \mathbf{L}_q \mathbf{D}_q^T \operatorname{vec}\left(\boldsymbol{\beta}^T \boldsymbol{\Delta}^{-1} \boldsymbol{\beta}\right)+\mathbf{L}_q \boldsymbol{\tau}_0+\left(\bm{g}_{i\bm{y}}^T \otimes \mathbf{I}_q\right) \operatorname{vec}\left(\mathbf{L}_q \bA\bC_2\right), \\
\bm\eta_{3} & = \operatorname{vech}(\bm\Delta^{-1}), \\
\bm{\eta}_{4} & = \operatorname{vech}(\bm\Delta^{-1}\bm\beta),\\
\bm\eta_{i\bm{y}5} & = -\frac{1}{2} \mathbf{J}_q \mathbf{D}_q^T \operatorname{vec}\left(\boldsymbol{\beta}^\T \boldsymbol{\Delta}^{-1} \boldsymbol{\beta}\right)+\mathbf{J}_q \boldsymbol{\tau}_0+\left(\bm{g}_{i\bm{y}}^\T \otimes \mathbf{I}_{k_q}\right) \operatorname{vec}\left(\mathbf{J}_q \bA\bC_2\right).
\end{aligned}
$$
The sufficient statistic is 
$$
\mathbf{T}(\bX_i, \bW_i) = \begin{pmatrix}
\operatorname{vec}(\bX_i^\T) \\
\bW_i \\
\mathbf{D}_p^\T \mathbf{D}_p \operatorname{vec}(\bX_i^\T \bX_i) \\
\operatorname{vec}\left(\sum_{j=1}^{m_i}\bX_{ij} \bW_i^\T\right)\\
\mathbf{J}_q \operatorname{vech}\left(\bW_i \bW_i^\T\right)
\end{pmatrix}
$$
Using Theorem 2.1 in \citet{bura2015sufficient} and the assumption that $E(\boldf_{ijy}) = E(\bm{g}_{i\bm{y}}) = \boldsymbol{0}$, the $i$th cluster-specific central subspace is given by the subspace spanned by
$$
\begin{aligned}
\bm{\eta}_{i\bm{y}} - {E}(\bm\eta_{i\bm{y}}) & = \begin{pmatrix}
\left(\mathbf{F}_{yi} \otimes \bI_p \right)\operatorname{vec}\left(\bm\Delta^{-1}\bGamma_i \bC_1 \right)\\
\left(\bm{g}_{i\bm{y}}^T \otimes \mathbf{I}_q\right) \operatorname{vec}\left(\mathbf{L}_q \bA\bC_2\right) - \bm\beta^\T\bm\Delta^{-1}\bGamma_i\bC_1\sum_{j=1}^{m_i}\boldf_{ijy}\\
\boldsymbol{0} \\
\boldsymbol{0} \\
\left(\bm{g}_{i\bm{y}}^\T \otimes \mathbf{I}_{k_q}\right) \operatorname{vec}\left(\mathbf{J}_q \bA\bC_2\right)
\end{pmatrix} 
\\[1em] & = \begin{pmatrix}
\left(\bI_{m_i} \otimes \bm\Delta^{-1}\bGamma_i \bC_1\right) \operatorname{vec}\left(\mathbf{F}_{yi}^\T\right)  \\[.5em]
\mathbf{L}_q \bA\bC_2 \bm{g}_{i\bm{y}} - \left(\boldsymbol{1}_{m_i} \otimes \bm\beta^\T \bm\Delta^{-1}\bGamma_i\bC_1 \right) \operatorname{vec}\left(\mathbf{F}_{yi}^\T\right)  \\
\boldsymbol{0} \\
\boldsymbol{0} \\
\mathbf{J}_q \bA\bC_2 \bm{g}_{i\bm{y}}
\end{pmatrix} \\
\end{aligned}
$$
Since the third and fourth elements of the last expressions are zero, then the minimal sufficient statistic is 
$$
\mathbf{T}(\bX_i, \bW_i) = \begin{pmatrix}
\operatorname{vec}(\bX_i^\T) \\
\bW_i \\
\mathbf{J}_q \operatorname{vech}\left(\bW_i \bW_i^\T\right)
\end{pmatrix}
$$
with the corresponding cluster specific central subspace being the one spanned by
$$
\begin{pmatrix}
\left(\bI_{m_i} \otimes \bm\Delta^{-1}\bGamma_i \bC_1\right) \operatorname{vec}\left(\mathbf{F}_{yi}^\T\right)  \\[.5em]
\mathbf{L}_q \bA\bC_2 \bm{g}_{i\bm{y}} - \left(\boldsymbol{1}_{m_i}^\T \otimes \bm\beta^\T \bm\Delta^{-1}\bGamma_i\bC_1 \right) \operatorname{vec}\left(\mathbf{F}_{yi}^\T\right)  \\
\mathbf{J}_q \bA\bC_2 \bm{g}_{i\bm{y}}
\end{pmatrix} \\
= \begin{pmatrix}
\bI_{m_i} \otimes \bm\Delta^{-1}\bGamma_i \bC_1 & \boldsymbol{0} \\
-\boldsymbol{1}_{m_i}^\T \otimes \bm\beta^\T \bm\Delta^{-1}\bGamma_i\bC_1  & \mathbf{L}_q \bA\bC_2 \\
\boldsymbol{0} & \mathbf{J}_q \bA\bC_2
\end{pmatrix} \begin{pmatrix}
\operatorname{vec}\left(\mathbf{F}_{yi}^\T\right)
\\ \bm{g}_{i\bm{y}},
\end{pmatrix}
$$ 
which is the same as those spanned by 
\[
\bm{\Theta}_i = \begin{pmatrix}
\bI_{m_i} \otimes \bm\Delta^{-1}\bGamma_i 
& \boldsymbol{0} \\
- \boldsymbol{1}_{m_i}^\T \otimes \bm\beta^\T \bm\Delta^{-1}\bGamma_i & \mathbf{0} \\
\boldsymbol{0} &  \bA
\end{pmatrix},
\]
as written in the statement of the Proposition.

\subsection{Proof of Proposition 6.2}
Similar to the proof of Proposition 6.1, the given model implies that conditional on the random effects $(\bm\Gamma_i, \bA_i)$, the joint distribution of $(\bX_{ij}, \bW_{ij}) \mid y_{ij}$ belong to an exponential family 
\[
f(\bX_{ij}, \bW_{ij} \mid y_{ij}) \propto \exp \left\{\mathbf{T}^T(\bX_{ij}, \bW_{ij}) \boldsymbol{\eta}_{ijy}-\psi\left(\boldsymbol{\eta}_{ijy}\right)\right\}
\]
with $\mathbf{T}^T(\bX_{ij}, \bW_{ij}) = (\bX_{ij}^\top - \bm\mu, \bW_{ij} - \bmu_W)^\top$ a sufficient statistic,  $$\bm\eta_{ijy} = \left( \bm\Delta^{-1} \bm\Gamma_{i}\bC \left\{\boldf_{ijy} - E(\boldf_{ijy}) \right\},  \left(\bA_i - \bm\beta^\top \bm\Delta^{-1} \bm\Gamma_{i}\bC \right) \left\{\boldf_{ijy} - E(\boldf_{ijy}) \right\} \right)$$ 
 a natural parameter and $\psi\left(\boldsymbol{\eta}_{ijy}\right)$ an appropriate (conditional) cumulant function. In this model, \citet{bura2015sufficient} shows that $\bTheta_i^\top   \mathbf{T}(\bX_{ij}, \bW_{ij})$ with $[\bTheta_i] = [\bm\Delta^{-1}\bm\Gamma_i\bC, -\bm\beta^\T\bm\Delta^{-1}\bm\Gamma_i\bC + \bA_i] $ is the cluster-specific minimal sufficient reduction for the regression of $y_{ij}$ on both $\bX_{ij}$ and $\bW_{ij}$. 
}

\section{Details and simulation results for random effects SDR with continuous predictors}

\subsection{Details of the MCEM algorithm for the RPFC model}
In the second stage of the estimation procedure, we apply the MCEM algorithm iterating between the following two-steps:

\underline{E-step}:
Let $\bm{\Psi}^{(0)}$ denote the estimates at the current iteration of the MCEM algorithm. Then we define the Q-function as $Q(\bm{\Psi}; \bm{\Psi}^{(0)}) = \sum_{i=1}^n \int \ell_{ci}(\bm{\Psi}) p(\bV_i;\bZ_{i}, \by_{i}, \bm{\Psi}^{(0)}) \, d \bV_i$, where \\ $p(\bV_i; \bZ_{i}, \by_{i}, \bm{\Psi}^{(0)})$ generically denotes the conditional distribution of the random effects given the observed data and current estimates. Since this expectation does not possess a closed form, then we utilize Monte-Carlo integration instead to approximate this. Suppose we sample $T$ values $\mathbf{V}^{t} \sim \text{MN}(\bm{0}, \bm\Sigma^{(0)}, \mathbf{I}_d); t= 1,\ldots, T$; in all the simulations and application in the article, we set $T=400$. Noting that when $\bm\Sigma^{(0)} \widehat{\bGamma}_0 = \boldsymbol{0}$, all the sampled $\bV^{t}$ belong to the tangent space $T_{[\widehat{\bGamma}_0]}\Gr(p,d)$, then by defining $\bm\Gamma^{t} = \operatorname{Exp}_{[\widehat{\bm\Gamma}_0]}(\bV^{t})$ we can construct the weights
    $
    \widetilde{w}^{(0)}_{it} =  \exp[-2^{-1}m_i \log \vert \bm\Delta^{(0)} \vert - 2^{-1}p \log \vert \bL_i \vert  - 2^{-1} \operatorname{tr}\{\bL_i^{-1} (\bZ_{ij} - \bm\Gamma^t \bC^{(0)} \bH_{i})^\top (\bm\Delta^{(0)})^{-1} (\bZ_{i} - \bm\Gamma^t \bC^{(0)} \bH_{iy} )\}]$,
and $w^{(0)}_{it} = (\sum_{i=1}\widetilde{w}^{(0)}_{it})^{-1}\widetilde{w}^{(0)}_{it}$, such that the $w_{it}^{(0)}$'s are normalized to sum to one for each cluster. The Q-function is then approximated as
\begin{align*}
 Q(\bm{\Psi} ; \bm{\Psi}^{(0)}) \approx \sum_{i=1}^{n} \sum_{t=1}^{T} w^{(0)}_{it} \left[\log \left\{\text{MN}(\bZ_{i}; \bm\Gamma^t\bC\bH_{iy}, \bm\Delta, \bL_i) \right\} + \log \left\{\text{MN}(\bV^t; \mathbf{0}, \bm\Sigma, \mathbf{I}_d)\right\}\right]. 
\end{align*}

\underline{M-step}:
We update the remaining parameters in the RPFC model as $\bm{\Psi}^{(1)} = \arg\max_{\bm{\Psi}} Q(\bm{\Psi}; \bm{\Psi}^{(0)})$, and achieve this via a series of conditional updates. First, the update for $\bm\Delta$ can be straightforwardly obtained in closed-form   
    $$
   \bm\Delta^{(1)} = \dfrac{1}{N-n} \sum_{i=1}^{n} \sum_{t=1}^{T} w^{(0)}_{it} (\bZ_{i} - \bm\Gamma^t \bC^{(0)}\bH_{iy})\bL_i^{-1}(\bZ_{ij} - \bm\Gamma^t \bC^{(0)}\bH_{iy})^\top.
    $$
Next, conditional on $\bm\Delta^{(1)}$ and setting the derivative of the approximated Q-function with respect to $\bC$ equal to zero, we have 
    $$    \sum_{i=1}^{n} \sum_{t=1}^{T}w^{(0)}_{it}\bm\Gamma^{t\top} \left(\bm\Delta^{(1)} \right)^{-1} \bm\Gamma^t \bC \mathbf{F}_{iy} =  \left\{ \sum_{i=1}^{n}\sum_{t=1}^{T}  w^{(0)}_{it} \bm\Gamma^{t\top}\left(\bm\Delta^{(0)} \right)^{-1}\mathbf{G}_{iy} \right\},  
    $$
    where $\mathbf{F}_{iy} = \bH_{iy}\bL_i^{-1}\bH_{iy}^\top$ and $\mathbf{G}_{iy}  = \bZ_i \bL_i^{-1} \bH_{iy}^\top$. Noting this equation has the form \\ $\sum_{i=1}^{n} \sum_{t=1}^{T} \mathbf{S}_{it} \bC \mathbf{F}_{iy} = \mathbf{E}$, where $\mathbf{S}_{it} = w^{(0)}_{it} \bm\Gamma^{t\top} \left(\bm\Delta^{(0)} \right)^{-1} \bm{\Gamma}^t$ and \\ $\mathbf{E} = \sum_{i=1}^{n}\sum_{t=1}^{T} \allowbreak  w^{(0)}_{it} \bm\Gamma^{t\top}\left(\bm\Delta^{(0)} \right)^{-1}\mathbf{G}_{iy}^\top$, then we can equivalently write $\bm{\Xi} \text{vec}(\bC) = \text{vec}(\mathbf{E})$ where $\bm{\Xi} = \sum_{i=1}^{n}\sum_{t=1}^{T}\mathbf{F}_{iy} \otimes \mathbf{S}_{it}$. It follows that 
    $\text{vec}(\bC^{(1)}) = \bm{\Xi}^{-1} \text{vec}(\mathbf{E})$, and a closed-form update is obtained.
    
    Finally,
    if $\bm\Sigma$ is not assumed to have additional structure besides being orthogonal to $\widehat{\bm\Gamma}_0$, then it is straightforward algebra to show that maximizing the approximated Q-function leads to the update
    $\bm\Sigma^{(1)}  = \sum_{i=1}^{n} \sum_{t=1}^{T} w^{(0)}_{it} \bV^{t} (\bV^{t})^\top$. This form of update ensures $\bm\Sigma^{(1)}$ is orthogonal to $\widehat{\bGamma}_0$ estimated from the first stage. Otherwise, if $\bm\Sigma$ is structured and characterized by a set of parameters denoted here as $\Phi$, then we can update these parameters correspondingly.
    For instance, in the simulations in Section 5 of the main text we consider settings where $\bm\Sigma = \sigma^2 \mathbf{K}$ and $\bK = \bI_p - \bGamma_0 \bGamma_0^\top$. Then it is straightforward to show that the update for $\sigma^2$ is given by
    $
    (\sigma^2)^{(1)} = \{d(p-d)\}^{-1}\sum_{t=1}^{T} \text{tr}(w^{(0)}_{it} \widehat{\bK}  \bV^t \bV^{t^\top}) = \{d(p-d)\}^{-1}\sum_{t=1}^{T} \text{tr}\{w^{(0)}_{it} \left(\bI_p - \widehat{\bGamma}_0 \widehat{\bGamma}_0^\top\right)  \bV^t \bV^{t^\top}\}.
    $

We iterate between the above two steps of the MCEM algorithm until convergence e.g., successive changes in the approximated marginal log-likelihood function, $\ell(\bm{\Psi} \vert \widehat{\bGamma}_0) = \sum_{i=1}^{n} \ell_i(\bm\Psi \vert \widehat{\bGamma}_0)$, where $\ell_{i}(\bm\Psi \vert \widehat{\bGamma}_0 ) = 
\log (\int \text{MN}(\bZ_{i};\bm\Gamma_i \bC \bH_{iy}, \bm\Delta, \bL_i) \, \text{MN}(\mathbf{V}_i; \mathbf{0}, \bm\Sigma, \mathbf{I}_d)  d\mathbf{V}_i)$ for $i=1,\ldots, n$, are smaller than some certain tolerance value.
Note $\ell(\bm{\Psi} |\widehat{\bGamma}_0)$ can be approximated using the proposed normalized weights in the MCEM algorithm, $\ell(\bm{\Psi}^{(0)}\vert \widehat{\bGamma}_0 ) \approx \sum_{i=1}^{n} \sum_{t=1}^{T} w^{(0)}_{it}$. 

Let $\widehat{\bm\Psi} =  \{\hat{\bC}^\top, \text{vech}(\hat{\bm\Delta})^\top, \text{vech}(\hat{\bm\Sigma})^\top\}^\top$  denote the estimator of $\bm\Psi$ upon convergence of the two-stage estimation procedure. Then the estimate of the overall fixed effect central subspace is given by $[\hat{\bm\Theta}_0] = \hat{\bm\Delta}^{-1}[\hat{\bm\Gamma}_0]$. Furthermore, we can predict the cluster-specific random effect central subspaces as follows:
For cluster $i$, we first compute a prediction of $\mathbf{V}_i$ on the tangent space $T_{[\bm\Gamma_0]}\text{Gr}(p, d)$ as the mean of the conditional distribution $\widehat{\bV}_i = \int \mathbf{V}_i p(\mathbf{V}_i; \bX_{ij}, \bm{y}_{i}, \widehat{\bm{\Psi}}) \approx \sum_{t=1}^T w^{(\infty)}_{it} \mathbf{V}^t$, where $\mathbf{V}^t \sim \text{MN}(\mathbf{0}, \widehat{\bm\Sigma}, \mathbf{I}_d)$ and $w^{(\infty)}_{it}$ denotes the normalized weights $w^{(0)}_{it}$ of the MECM algorithm evaluated at $\widehat{\bm{\Psi}}$. 
We then obtain a prediction as $\bm{\widehat\Gamma}_i = \Exp_{[\widehat{\bm\Gamma}_0]}(\widetilde{\bm{V}}_i)$ and $[\widehat{\bm\Theta}_i] = \widehat{\bm\Delta}^{-1}[\widehat{\bm\Gamma}_i]$, for $i=1,\ldots, n$. 
While it is possible to use other predictors e.g., the mode or median of $p(\mathbf{V}_i; \bX_{ij}, \bm{y}_{i}, \widehat{\bm{\Psi}})$, in our empirical study we experimented with several choices and found that using the mean of the conditional distribution tended to be the most accurate.

\subsection{Further simulation results}
When the random effect covariance matrix $\bm\Sigma$ is (known to be) isotropic in structure, Table \ref{tab:inversemodels_p7-isotropic} demonstrates that the RPFC model had the overall best performance. For estimating the overall fixed effect central subspace $[\bm\Theta_0]$, SPFC had the poorest performance, while when the clusters relative homogeneous (e.g., $\sigma^2 = 0.04$) RPFC still produced a lower estimation error than GPFC. This latter result is interesting because while both the RPFC and GPFC models use the same estimator for $\bm\Gamma_0$, the former (again) performed better due to its superior performance at recovering the residual covariance matrix $\bm\Delta$. When $\sigma^2 = 0.50$, RPFC had worse performance than GPFC and SPFC when $n=100$ and $n=500$, but its performance greatly improved with increasing $n$ and tended to be similar to GPFC for $n=1000$.

Turning to estimation of $\bm\Sigma$, RPFC consistently outperformed SPFC, and its estimation error decreased with increasing $n$ while this does not occur for SPFC. 

Finally, for predicting the cluster-specific random effect central subspaces, RPFC consistently outperformed SPFC especially when $\sigma^2$ was small. This reflected the former's capacity to borrow strength across clusters in the SDR operation, which in turn improved the overall performance at predicting the $[\bm\Theta_i]$'s. When the amount of heterogeneity between clusters increased, RPFC was still able to predict $[\bm\Theta_i]$ better than SPFC, although the differences between the two methods became smaller. Note the performance relating to prediction of cluster-specific random effect central subspaces does not tend to decrease substantially when the number of clusters $n$ increases: this is a consequence of the cluster sizes $m_i$ being bounded as $n$ grows in our simulation design.  

\begin{table}[H]
\centering
\caption{Simulation results for random effects SDR where an isotropic random effects covariance matrix $\bm\Sigma = \sigma^2 \bK$ is used. The methods compared include the RPFC, GPFC, and SPFC models. Performance is assessed in terms of estimating the overall fixed effect central subspace $\bm\Delta^{-1}[\bGamma_0]$, the random effects covariance matrix $\tilde{\bm\Sigma}$, and predicting the cluster-specific random effect central subspaces $[\bm\Theta_i]$. For each measure, the method/s with the lowest average Frobenius error in each row is highlighted.}
\resizebox{.9\textwidth}{!}{\begin{tabular}{lcl >{\bfseries}ccc >{\bfseries}ccH >{\bfseries}c c}
  \toprule[1.5pt]
   & & & \multicolumn{3}{c}{$\bm\Delta^{-1}[\bm\Gamma_0]$} & \multicolumn{3}{c}{${\bm\Sigma}$} & \multicolumn{2}{c}{$[\bm\Theta_i]$} \\ 
$\sigma^2$ & Model & $n$  & \normalfont{RPFC} & GPFC & SPFC & \normalfont{RPFC} & SPFC & GPFC & \normalfont{RPFC} & SPFC \\ 
\cmidrule(lr){4-6} \cmidrule(lr){7-9} \cmidrule(lr){10-11}
\addlinespace
0.04 & M1 & 100 & 0.29 (0.09) & 0.34 (0.11) & 1.23 (0.23) & 0.01 (0.01) & 0.40 (0.06) & 0.10 (0.00) & 0.39 (0.02) & 1.14 (0.04)\\
 &  & 500 & 0.18 (0.05) & 0.27 (0.07) & 1.11 (0.24) & 0.01 (0.00) & 0.40 (0.06) & 0.10 (0.00) & 0.39 (0.01) & 1.14 (0.04)\\
 &  & 1000 & 0.15 (0.04) & 0.28 (0.08) & 1.14 (0.22) & 0.01 (0.00) & 0.40 (0.06) & 0.10 (0.00) & 0.39 (0.01) & 1.14 (0.04)\\
\addlinespace
 & M2 & 100 & 0.86 (0.29) & 0.79 (0.25) & 2.46 (0.32) & 0.02 (0.01) & 0.46 (0.04) & 0.09 (0.00) & 0.78 (0.08) & 1.52 (0.05)\\
 &  & 500 & 0.48 (0.18) & 0.52 (0.14) & 2.46 (0.34) & 0.02 (0.01) & 0.47 (0.05) & 0.09 (0.00) & 0.70 (0.03) & 1.52 (0.04)\\
 &  & 1000 & 0.37 (0.12) & 0.45 (0.12) & 2.44 (0.34) & 0.02 (0.01) & 0.45 (0.04) & 0.09 (0.00) & 0.69 (0.02) & 1.51 (0.04)\\
\addlinespace
0.1 & M1 & 100 & 0.42 (0.14) & 0.61 (0.17) & 1.35 (0.24) & 0.04 (0.01) & 0.32 (0.05) & 0.24 (0.00) & 0.49 (0.02) & 1.17 (0.03)\\
 &  & 500 & 0.19 (0.06) & 0.56 (0.13) & 1.23 (0.21) & 0.04 (0.01) & 0.32 (0.05) & 0.24 (0.00) & 0.49 (0.01) & 1.16 (0.02)\\
 &  & 1000 & 0.14 (0.04) & 0.55 (0.14) & 1.20 (0.20) & 0.04 (0.01) & 0.32 (0.05) & 0.24 (0.00) & 0.49 (0.01) & 1.16 (0.02)\\
\addlinespace
 & M2 & 100 & 1.29 (0.44) & 1.25 (0.35) & 2.47 (0.31) & 0.08 (0.02) & 0.39 (0.04) & 0.22 (0.00) & 1.07 (0.09) & 1.56 (0.03)\\
 &  & 500 & 0.69 (0.30) & 0.89 (0.20) & 2.48 (0.34) & 0.05 (0.01) & 0.40 (0.04) & 0.22 (0.00) & 0.98 (0.05) & 1.55 (0.03)\\
 &  & 1000 & 0.48 (0.20) & 0.82 (0.14) & 2.47 (0.35) & 0.04 (0.01) & 0.40 (0.04) & 0.22 (0.00) & 0.96 (0.02) & 1.55 (0.02)\\
\addlinespace
0.5 & M1 & 100 & \normalfont  2.13 (0.44) & \bf 1.80 (0.34) & 1.84 (0.30) & 0.60 (0.08) & 0.81 (0.03) & 1.22 (0.00) & 0.62 (0.03) & 1.21 (0.02)\\
 &  & 500 & \normalfont 1.87 (0.52) & \bf 1.67 (0.36) & 1.75 (0.30) & 0.54 (0.10) & 0.80 (0.03) & 1.22 (0.00) & 0.62 (0.01) & 1.20 (0.01)\\
 &  & 1000 & 1.57 (0.59) &  \bf 1.57 (0.36) & 1.69 (0.30) & 0.48 (0.11) & 0.80 (0.03) & 1.22 (0.00) & 0.62 (0.01) & 1.21 (0.01)\\
\addlinespace
 & M2 & 100 & \normalfont 2.65 (0.33) & \bf 2.34 (0.30) & 2.49 (0.32) & 0.79 (0.09) & 0.83 (0.04) & 1.12 (0.00) & 1.39 (0.02) & 1.60 (0.02)\\
 &  & 500 & \normalfont  2.31 (0.37) & \bf 2.21 (0.31) & 2.48 (0.31) & 0.71 (0.08) & 0.83 (0.04) & 1.12 (0.00) & 1.39 (0.01) & 1.60 (0.01)\\
 &  & 1000 & 2.16 (0.38) & 2.17 (0.31) & 2.49 (0.32) & 0.66 (0.10) & 0.83 (0.04) & 1.12 (0.00) & 1.39 (0.01) & 1.60 (0.01)\\
   \bottomrule[1.5pt]
   \label{tab:inversemodels_p7-isotropic}
\end{tabular}}
\end{table}

\FloatBarrier

\section{Selecting the structural dimension for random effects SDR} \label{sec:selectingdimension}
In this section, we focus on choosing $d$ in the setting of the RPFC model i.e., random effects sufficient dimension reduction with continuous predictors.

For the standard PFC model without random effects, \citet{cook2008} proposed to select the structural dimension $d$ by either a likelihood ratio test or via an information criteria. In this article, we adopt the latter approach when it comes to selecting $d$ for RPFC \citep[see also][for examples of where information criteria are employed to select the structural dimension in SDR]{ma2015validated,luo2021order}. In particular, 
we propose two computationally efficient approaches that allow for the selection of $d$ to be made \emph{prior} to fitting the (second stage of the) RPFC model.

In the first approach, since the structural dimension is equal to the dimension of $[\widehat{\bm\Gamma}_0]$, which is estimated via a GPFC model in the first stage of the estimation procedure, then we consider formulating an information criterion directly from this. Let $w \in \{0, 1, \ldots, \text{min}(r, p)\}$ be a candidate for $d$. Then after applying GPFC model with this candidate choice, let $\ell_g(w) = \ell_g(w; \widehat{\bm\mu}^\GPFC, \widehat{\bm\Gamma}^\GPFC, \widehat{\bm\beta}^\GPFC, \widehat{\bm\Delta}^\GPFC)$ denote the corresponding value of the maximized log-likelihood function, noting that the corresponding number of parameters involved is $h(w) = p(p+3)/2 + rw  + w(p-w)$. It follows that we can construct an information criterion of the form $-2\ell_g(w) + \rho h(w)$ where we consider the model complexity parameter as $\rho = 2$ or $\rho = \log(N)$ corresponding to the global Akaike information criterion (GAIC) and the global Bayesian information criterion (GBIC) respectively. We select $d$ that minimizes either GAIC or GBIC.

In the second approach, by noting the structural dimension is the same across all clusters in the RPFC model, we consider selecting $d$ via the SPFC model. Specifically, after fitting SPFC with a dimension candidate $w$, let $\ell_i(w; \widehat{\bm\mu}^{\text{SPFC}}_i, \widehat{\bm\Gamma}^{\text{SPFC}}_i, \widehat{\bm\beta}^{\text{SPFC}}_i, \widehat{\bm\Delta}^{\text{SPFC}}_i)$ denote the corresponding value of the maximized log-likelihood function for the $i$-th cluster. Then an information criterion for the $i$-th cluster can be defined as $-2\ell_i(w) + \rho_i h(w)$, and we select $d$ by minimizing $\sum_{i=1}^{n} \left\{-2\ell_i(w) + \rho_i h(w)\right\}$. As in the first approach, we consider a separate Akaike information criterion (SAIC) by setting all $\rho_i=2$, and a separate Bayesian information criterion (SBIC) by setting all  $\rho_i = \log(m_i)$.

Note a big advatange of adopting the above methods to select $d$ is computational efficiency: instead of fitting the RPFC model to each candidate $d$ using the two-stage estimation procedure, we can select $d$ using either the GPFC and SPFC models, which are computationally very scalable to fit. 

\begin{figure}[tb]
    \centering
    \includegraphics[width = 0.9\textwidth]{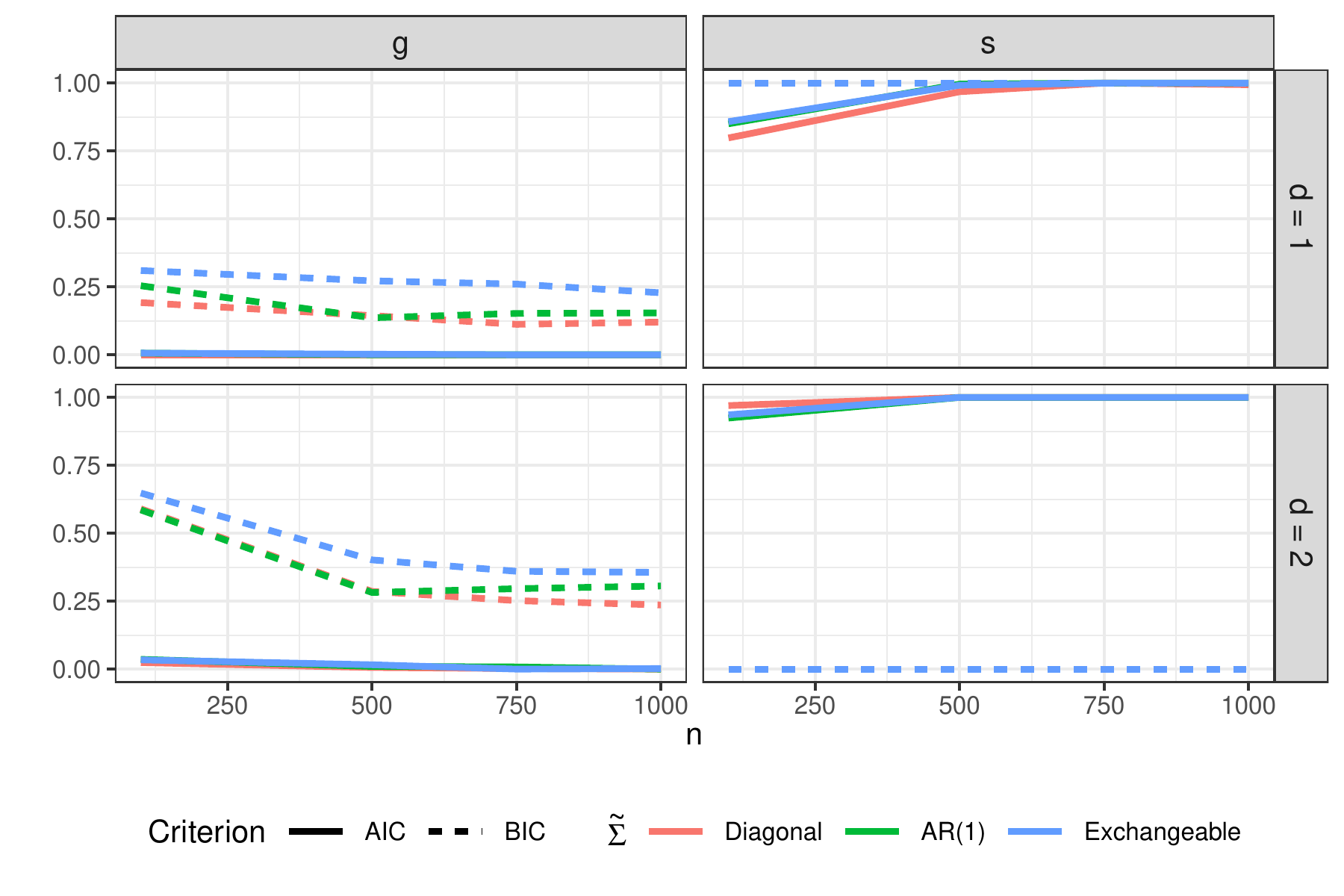}
    \caption{Proportion of 500 simulated datasets where the proposed information criteria methods selected the correct structural dimension for the RPFC model. The top and bottom panels correspond to models M1 and M2, whose true $d$ equals 1 and 2 respectively.}
    \label{fig:choice-of-d}
\end{figure}

To assess the finite sample performance of the above procedure, we conduct a simulation study where clustered data were generated from the two inverse regression models as in Section 5 of the main text.
Recall the true structural dimension of models M1 and M2 are $d=1$ and $d=2$, respectively. Figure \ref{fig:choice-of-d} shows that SAIC exhibits the best performance among the four criterion considered for selecting $d$, when candidates values of $d$ ranging from 1 to 5 are considered. It is also the only method whose performance tended to improve when the number of clusters $n$ increases. 
The two information criteria derived from the GPFC model exhibited poor performance, possibly because $\textsc{GAIC}$ and $\textsc{GBIC}$ are constructed from a model that ignores the heterogeneity among clusters and thus effectively underfits the data. Finally, the $\textsc{SBIC}$ consistently chose $\hat{d}=1$, meaning its model complexity penalty was likely too severe.  We leave theoretical investigation of the proposed information criteria for selecting $d$ in the RPFC model as an avenue for future research. 

\revise{Finally, we note the SAIC and SBIC methods can still be applied to select the structural dimension $d$ for continuous covariates in the RMIR models with time-invariant binary covariates in the main text.
We use this approach to select $d$ for the data application in Section 7 of the main paper.}

\FloatBarrier

\revise{
\section{Estimation details for RMIR models with mixed continuous and binary predictors}
\label{subsection:estimation_extension}
\subsection{Models for binary predictors}
For both examples of RMIR defined in Section 6.1 and 6.2 of the main text, the corresponding parameters can be estimated by maximizing two separate likelihood functions, one from the conditional distribution $\bX_{i} \mid (\by_{i}, \bW_i)$, and another from the conditional distribution $\bW_{i} \mid \by_i$. With a slight abuse of notation, for the case of time-invariant binary covariates, we let $\bW_{ij} = \bW_i$ for all $j=1,\ldots, m_i$. We address the maximization of these two components separately. 

First, for the Ising model of $\bW_{i} \mid \by_i$ in the model with time-invariant binary covariates (model 
(8) in the main paper), we use the same pseudo-conditional likelihood approach as in \citet{bura2023sufficient} to estimate parameters $\bm\tau_0$, $\bB$ and $\bC_2$. Recall that $\bm\tau_0$ and $\bB$ has $q(q+1)/2$ rows, each row corresponding to one pair $(k, k^\prime)$ for $k, k^\prime=1,\ldots q$. Hence we let $\tau_{0, kk}$ and $\bm{a}_{kk}^\top$ denote the row of $\bm\tau_0$ and $\bB$ corresponding to the pair $(k, k^\prime)$ respectively.  The pseudo-conditional likelihood is obtained from the node-wise logistic regression models of one component $W_{ik}$ on the remaining components $\bW_{i, -k}$, the basis $\bm{g}_{i\bm{y}}$ and its interaction. That is, we estimate $\bm\tau_0$, $\bB$ and $\bC_2$ by maximizing $\sum_{i=1}^{n} \sum_{k=1}^{q} \left[W_{ik}\zeta_{ik} - \log(1 + e^{\zeta_{ik}})  \right]$ where
\[
\begin{aligned}
\zeta_{ik} = \log \frac{\operatorname{Pr}\left(W_{ik}=1 \mid \bW_{i,-k}, \bm{y}_i\right)}{\operatorname{Pr}\left(W_{ik}=0 \mid \bW_{i, -k}, \bm{y}_i \right)}=\tau_{0, kk}+ \bm{b}_{kk}^\top \bC_2
\bm{g}_{i\by }+\sum_{k \neq k^{\prime}} \tau_{kk^{\prime} 0} W_{ik^{\prime}}+\sum_{j<j^{\prime}} \bm{b}_{kk^{\prime}}^\top \bC_2\bm{g}_{i\by} W_{ik^{\prime}}. 
\end{aligned}
\]
Computationally, this pseudo-conditional log-likelihood function is maximized via an iteratively reweighted least square algorithm.   For the model with time-variant binary covariate (model (9) in the main paper), we fit $k$ separate logistic mixed effect models for each component $\left\{\bW_{ijk} \right\}$ on $\boldf_{ijy}$, for $i=1,\ldots,n, j=1,\ldots, m_i$. 

\subsection{Models for continuous predictors}
For the parameters involved in the conditional distribution $\bX_{i} \mid (\by_{i}, \bW_i)$, we employ a two-stage method similar to the procedure given in subsection 3.2 of the main text. Specifically, the fixed effects of $[\bGamma_0]$ can be estimated by a global reduced-rank regression model that ignores the clustering, i.e 
\begin{equation}
\bX_{ij} = \widetilde{\bm\mu} + \widetilde{\bm\Gamma} \widetilde{\bC} \boldf_{ijy} + \widetilde{\bm\beta}\bW_{ij} +\widetilde{\bm{\varepsilon}}_{ij}  ,  ~\bm\varepsilon_{ij} \sim N_p(\boldsymbol{0}, \widetilde{\bm\Delta}).
\label{eq:reducedrankmodel}
\end{equation}
The reduced-rank model \eqref{eq:reducedrankmodel} contains $\bX_{ij}$ as the response and two sets of covariates, one reduced-rank $\boldf_{ijy}$,  and another non-reduced-rank $\bW_{ij}$. Hence this model can be fitted using the estimation algorithm in \citet[Chapter 3]{reinsel2022multivariate} to maximize
\begin{equation}
\begin{aligned}
\ell_g(\widetilde{\bm\mu}, \widetilde{\bm\Gamma}, \widetilde{\bm\beta}, \widetilde{\bm\Delta}) = -\dfrac{N}{2} \log \vert \widetilde{\bm\Delta} \vert - \dfrac{1}{2} \sum_{i=1}^{n}\sum_{j=1}^{m_i} & \left(\bX_{ij} - \widetilde{\bm\mu} - \widetilde{\bm\Gamma} \widetilde{\bC} \boldf_{ijy} - \widetilde{\bm\beta} \bW_{ij} \right)^{\top} \widetilde{\bm\Delta}^{-1} \\ & \times \left(\bX_{ij} - \widetilde{\bm\mu} - \widetilde{\bm\Gamma} \widetilde{\bC} \boldf_{ijy} - \widetilde{\bm\beta} \bW_{ij} \right).
\end{aligned}
\label{eq:globalPFCllhextension}
\end{equation}
To establish the consistency of this approach in estimating $[\bGamma_0]$, we impose a technical condition in addition to Conditions C1 and C2 in the main text.
\setcounter{condition}{2}
\begin{condition}
Let $\widehat{\bm\Sigma}_{fw} = N^{-1}\sum_{i=1}^{n}\sum_{j=1}^     
     {m_i} \boldf_{ijy}(\bW_{ij} - \overline{\bW})^\top$ and $\widehat{\bm\Sigma}_{ww} = N^{-1}\sum_{i=1}^{n}\sum_{j=1}^{m_i} (\bW_{ij}  - \overline{\bW})(\bW_{ij} - \overline{\bW})^\top$, where $\overline{\bW} = N^{-1}\sum_{i=1}^{n}\sum_{j=1}^{m_i}\bW_{ij} $. As $N \to \infty$, the matrix 
     $
     \begin{pmatrix}
      \widehat{\bSigma}_{ff} & \widehat{\bSigma}_{fw} \\
      \widehat{\bSigma}_{wf} & \widehat{\bSigma}_{ww} \\
     \end{pmatrix}
     $ converges to a positive definite matrix 
     $
\begin{pmatrix}
      {\bSigma}_{ff} & {\bSigma}_{fw} \\
      {\bSigma}_{wf} & {\bSigma}_{ww} \\
     \end{pmatrix}.
$
\label{eq:condition1_extension}
\end{condition}
With this new additional condition, the following theorem justifies this approach for estimating $[\bm\Gamma_0]$ in this situation. 

\begin{theorem} \label{thm:RMIRfirststageconsistency}
Let $\hat{\bm\Gamma}_{0}$ denote the maximum likelihood estimate of $\tilde{\bm{\Gamma}}$ from the GPFC model i.e., obtained by maximizing equation \eqref{eq:globalPFCllhextension}. Assume Conditions C1-C2 in the main paper and condition \ref{eq:condition1_extension} are satisfied. Then $[\hat{\bm\Gamma}_0]$ $\xrightarrow{p} [\bm\Gamma_0^*]$ as $N \rightarrow \infty$.
\end{theorem}
After the first stage, the second stage contains an MCEM algorithm to estimate other parameters in the conditional distribution of $\bX_{ij} \mid (y_{ij}, \bW_{ij})$. Since they are similar to the algorithm given in subsection 3.2, we omitted the details.

\subsection{Proof of Theorem \ref{thm:RMIRfirststageconsistency}}

With this technical condition, the proof of this result is similar to that of Theorem 4.1 with the RPFC model, except with all the marginal covariance matrices replaced by the corresponding partial covariance matrices controlling for the non-reduced-rank covariates $\bW$. 

Using results for multivariate reduced rank regression \citep[Chapter 3, p.76][]{reinsel2022multivariate}, we obtain
$
[\widehat{\bC}] = [\widehat{\bm\Sigma}_{xx}^{1/2}\hat{\bm\Lambda}],
$
where 
\begin{itemize}
\item $\widehat{\bm\Lambda}$ is the $p \times d$ matrix formed of the first $d$ normalized eigenvectors of the matrix \\$\widehat{\bm\Sigma}_{xx}^{-1/2}\hat{\bm\Sigma}_{xf.w} \widehat{\bm{\Sigma}}_{ff.w}^{-1} \widehat{\bm\Sigma}_{fx.w} \widehat{\bm\Sigma}_{xx}^{-1/2}$, with
    \item $\widehat{\bm\Sigma}_{ff.w} = \widehat{\bSigma}_{ff} - \widehat{\bSigma}_{fw}\widehat{\bSigma}_{ww}^{-1}\widehat{\bSigma}_{wf}$,
    \item $\widehat{\bm\Sigma}_{xf.w} = \widehat{\bm\Sigma}_{fx.w}^\top = \widehat{\bSigma}_{xf} - \widehat{\bSigma}_{xw}\widehat{\bSigma}_{ww}^{-1}\widehat{\bSigma}_{wf}$.
\end{itemize}

By Condition \ref{eq:condition1_extension}, we have $\widehat{\bSigma}_{ff.w} \xrightarrow{p} \bSigma_{ff} - \bSigma_{fw} \bSigma_{ww}^{-1} \bSigma_{wf} = \bSigma_{ff.w}$ as $N\to\infty.$ By the law of large numbers and Condition \ref{eq:condition1_extension}. we have
\begin{equation}
\begin{aligned}
\widehat{\bm\Sigma}_{xw} & = \dfrac{1}{N} \sum_{i=1}^{n}\sum_{j=1}^{m_i} E\left\{(\bX_{ij} - \bar{\bX}_i)(\bW_{ij} - \overline{\bW})^\top\right\} + O_p\left(N^{-1/2} \right) 
\\ & = \dfrac{1}{N} \sum_{i=1}^{n}\sum_{j=1}^{m_i} E\left[\left\{\bm\Gamma_{i}\bC_1\boldf_{ijy} + \bm\beta(\bW_{ij} - \overline{\bW})\right\}(\bW_{ij} - \overline{\bW})^\top \right] + O_p\left(N^{-1/2} \right) \\
& = E\left(\bGamma_i \bC_1 \widehat{\bSigma}_{fw} \right) + \bm\beta \bSigma_{ww}  + O_p\left(N^{-1/2} \right).
\end{aligned}
\label{eq:lln_extension}
\end{equation}
Next, by the law of iterated expectation, we obtain
$$
\begin{aligned}
E\left(\bGamma_i \bC_1 \widehat{\bSigma}_{fw} \right) & = E\left[E\left(\bGamma_i \bC_1 \widehat{\bSigma}_{fw} \mid \bGamma_i \right)\right] & =  E \left( \bGamma_i   \right) \bSigma_{fw} \stackrel{(i)}{=} \bGamma_0 \mathbf{M} \bC_1 \bSigma_{fw} ,
\end{aligned}
$$
where step $(i)$ follows from the argument in the derivation of Lemma 1 with $\mathbf{M}$ as a (deterministic) non-singular $d\times d$ matrix. Substituting this result into equation~\eqref{eq:lln_extension}, as $ N \to \infty$,  we obtain
$$
\widehat{\bm\Sigma}_{xw}  = \bGamma_0 \mathbf{M} \bC_1 \bSigma_{fw} + \bm\beta \bSigma_{ww}  +O_p\left(N^{-1/2} \right).
$$
By a similar argument, as $N \to \infty$, we obtain
\begin{equation*}
\widehat{\bSigma}_{xf} = \bGamma_0 \mathbf{M} \bC_1 \bSigma_{ff} + \bm\beta \bSigma_{wf} +O_p\left(N^{-1/2} \right).  
\end{equation*}
Hence, as $N \to \infty$, we have 
$$
\begin{aligned}
\widehat{\bm\Sigma}_{xf.w} & = \bm\Gamma_0 \bM \bC_1 \bSigma_{ff} + \bm\beta \bSigma_{wf} - (\bGamma_0 \mathbf{M} \bC_1 \bSigma_{fw} + \bm\beta \bSigma_{ww}) \bSigma_{ww}^{-1} \bSigma_{wf} +O_p\left(N^{-1/2} \right)  \\  & =
\bm\Gamma_0 \bM \bC_1 \left(\bSigma_{ff} - \bSigma_{fw} \bSigma_{ww}^{-1} \bSigma_{wf}\right) = \bm\Gamma_0 \bM \bC_1 \bSigma_{ff.w} +O_p\left(N^{-1/2} \right).
\end{aligned}
$$
As $N \to \infty$ then, then under Condition C1 we obtain
$$
\widehat{\bm\Sigma}_{xf.w}\widehat{\bm\Sigma}_{ff.w}^{-1} \widehat{\bm\Sigma}_{fx.w} = \bGamma_0 \mathbf{M} \bC_1 \bm\Sigma_{ff.w} \bC_1^{\top}\mathbf{M}^\top \bGamma_0^\top +O_p\left(N^{-1/2} \right).
$$
Finally, we note $\bGamma_0 \mathbf{M} \bC_1 \bm\Sigma_{ff.w} \bC_1^{\top}\mathbf{M}^\top \bGamma_0^\top$ is a $p \times p $ matrix of rank $d$, so is the matrix \\ $\bm\Sigma_{xx}^{-1/2} \mathbf{M} \bC_1 \bm\Sigma_{ff.w} \bC_1^{\top}\mathbf{M}^\top \bGamma_0^\top \bm\Sigma_{xx}^{-1/2}$. Hence, as $N \to \infty$, the space spanned by its $d$ eigenvectors is equivalent to its column space. It follows that 
$$
[\widehat\bC] \stackrel{p}{\to} \bm\Sigma_{xx}^{1/2}[\bm\Sigma_{xx}^{-1/2} \bGamma_0 \mathbf{M} \bC_1\bm\Sigma_{ff.w} \bC_1^{\top}\mathbf{M}^\top \bGamma_0^\top \bm\Sigma_{xx}^{-1/2}] = [\bm\Gamma_0],
$$
as required.

\FloatBarrier
}

\revise{
\section{Additional results for application to longitudinal socioeconomic data}
This section provides some additional results for the application in Section 7 of the main text. 

\begin{figure}[H]
\centering
\includegraphics[width = \textwidth]{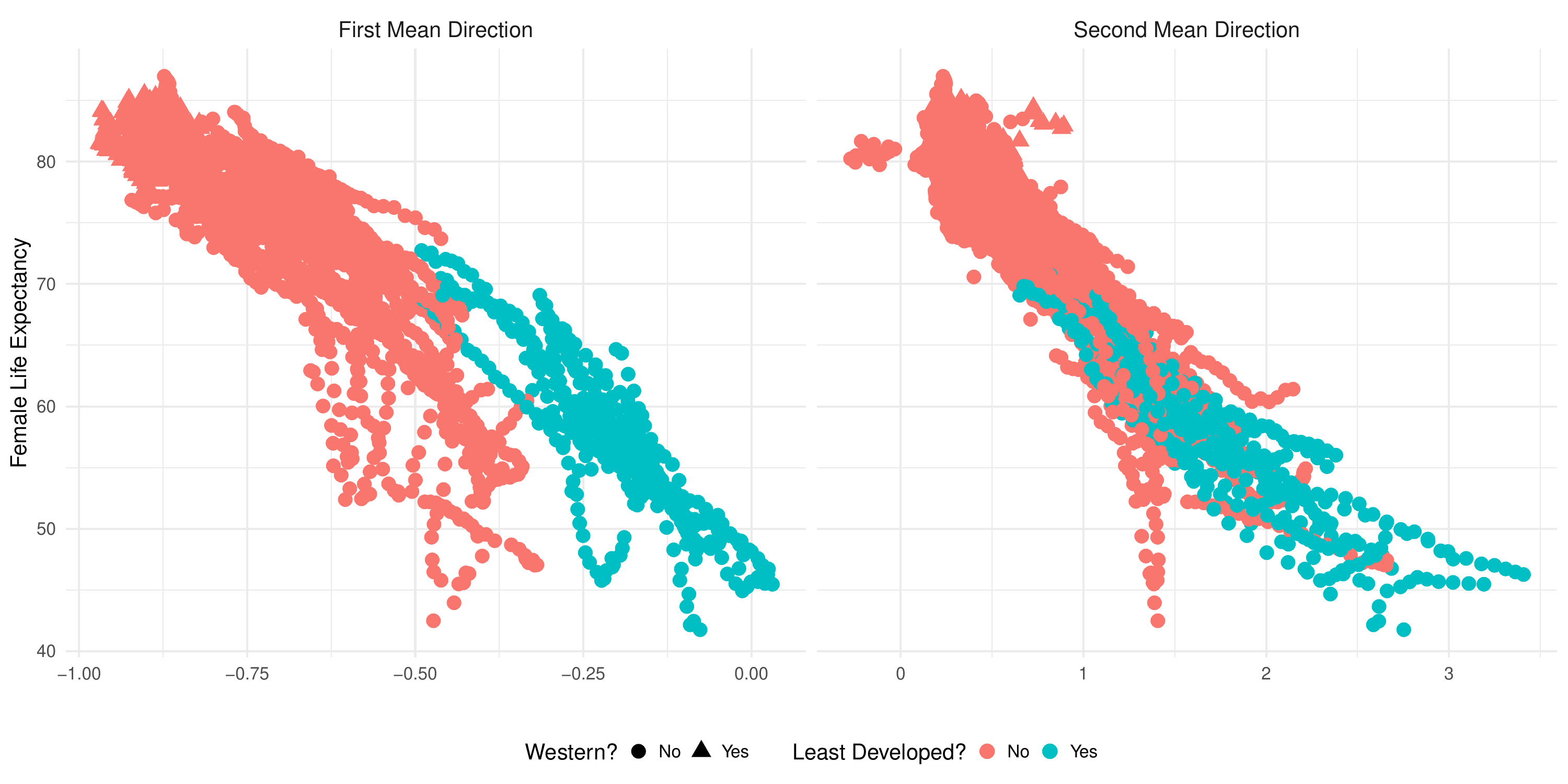}
\caption{Scatterplots of the female life expectancy across all countries and years $y_{ij}$ versus the estimated fixed effects sufficient predictors at the observation level i.e., $\bX_{ij}^\top \widehat{\bm\Delta}^{-1}\widehat{\bm\Gamma}_0 - \bW_{i}^\top\widehat{\bm\beta}^\top\widehat{\bm\Delta}^{-1}\widehat{\bm\Gamma}_0$, based on the RMIR model. Note the form of the fixed effects sufficient predictors at the observation level is formed by taking the component of sufficient predictor below Proposition 6.1 in the main text, and replacing $\bm{\Gamma}_i$ with $\bm{\Gamma}_0$. Different colors and shapes represent different combinations of the two binary covariates.}
\label{fig:centralsubspaces_meanDR_binary}
\end{figure}

\FloatBarrier

\begin{figure}[H]
\includegraphics[width = \textwidth, page = 1]{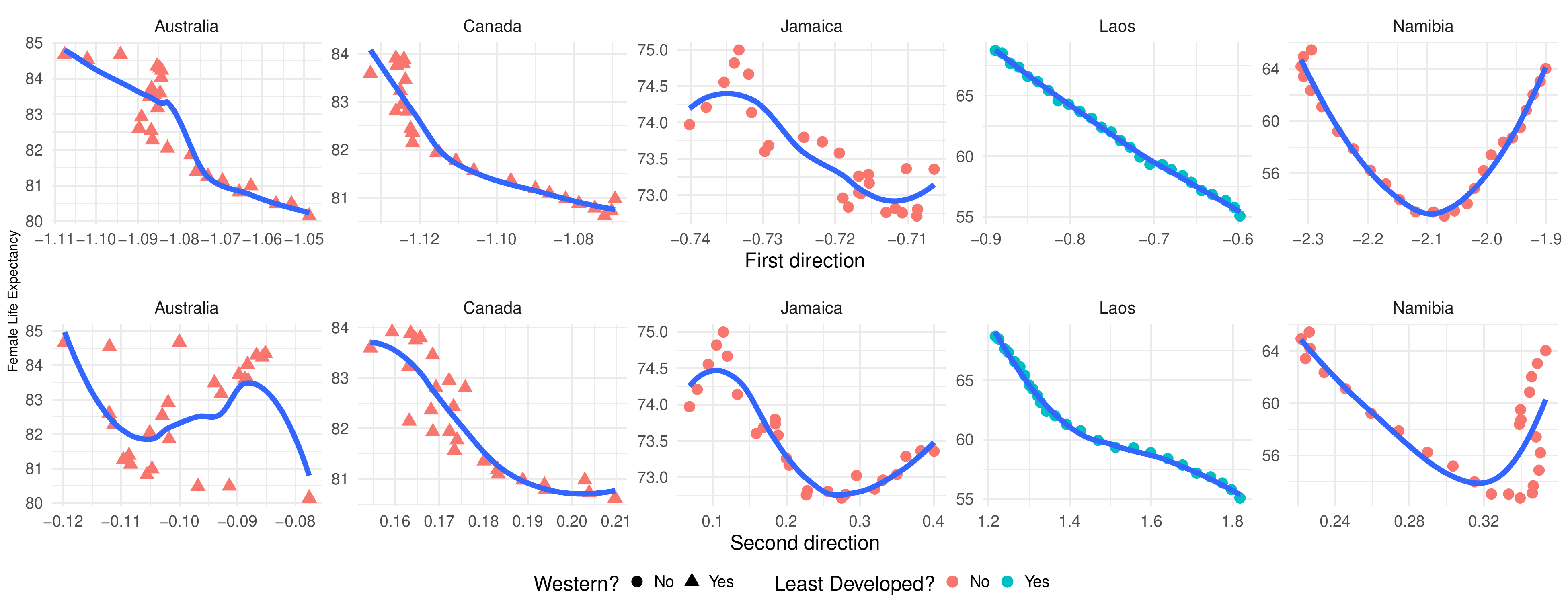}
\caption{Scatterplots of female life expectancy against estimated cluster-specific sufficient predictors at the observation level i.e., $\bX_{ij}^\top \widehat{\bm\Delta}^{-1}\widehat{\bm\Gamma}_i - \bW_{i}^\top\widehat{\bm\beta}^\top\widehat{\bm\Delta}^{-1}\widehat{\bm\Gamma}_i$ for select countries, based on the RMIR model. The blue line represents a corresponding LOESS curve smooth.\\\\ 
Results show that female life expectancy exhibits a strong, sometimes close to linear relationship with one or both of these sufficient predictors across many countries. However, in some countries such as Australia and Canada, the first sufficient predictor is more informative about the response than the second one. This is reversed though for other countries such as Oman and Jamaica, while in some other countries e.g., Laos, both sufficient predictors are strongly and close to linearly associated with female life expectancy. 
}
\label{eq:plot_response_suffpred}
\end{figure}
}

\FloatBarrier

\end{document}